\documentclass[prb,amsmath,twocolumn,superscriptaddress,showpacs]{revtex4}

\usepackage{amsfonts}
\usepackage{amsmath}
\usepackage{amssymb}
\usepackage{graphicx}
\usepackage{txfonts}
\usepackage{color}
\usepackage{hyperref}

\renewcommand{\vec}[1]{\mathbf{#1}}

\begin{document}
	\title{Dynamic response functions and helical gaps in interacting\\ Rashba nanowires with and without magnetic fields}
	\author{Christopher J. Pedder}
	\affiliation{Physics and Materials Science Research Unit, University of Luxembourg, L-1511 Luxembourg}
	\author{Tobias Meng}
	\affiliation{Institut f\"ur Theoretische Physik, Technische Universit\"at Dresden, 01062 Dresden, Germany}
	\author{Rakesh P. Tiwari}
	\affiliation{Department of Physics, University of Basel, Klingelbergstrasse 82, CH-4056 Basel, Switzerland}
	\author{Thomas L. Schmidt}
	\email{thomas.schmidt@uni.lu}
	\affiliation{Physics and Materials Science Research Unit, University of Luxembourg, L-1511 Luxembourg}
	
	\date{\today}
	
\begin{abstract}
A partially gapped spectrum due to the application of a magnetic field is one of the main probes of Rashba spin-orbit coupling in nanowires. Such a ``helical gap'' manifests itself in the linear conductance, as well as in dynamic response functions such as the spectral function, the structure factor, or the tunnelling density of states. In this paper, we investigate theoretically the signature of the helical gap in these observables with a particular focus on the interplay between Rashba spin-orbit coupling and electron-electron interactions. We show that in a quasi-one-dimensional wire, interactions can open a helical gap even without magnetic field. We calculate the dynamic response functions using bosonization, a renormalization group analysis, and the exact form factors of the emerging sine-Gordon model. For special interaction strengths, we verify our results by refermionization. We show how the two types of helical gaps, caused by magnetic fields or interactions, can be distinguished in experiments.
\end{abstract}
	
	
	\maketitle
	
\section{Introduction}

In the past few years, there has been a surge of interest in one-dimensional (1D) semiconductor wires with Rashba spin-orbit coupling (RSOC), largely as a result of their potential to host quasiparticles behaving as Majorana fermions. In proximity to a bulk $s$-wave superconductor and subjected to a magnetic field perpendicular to the Rashba axis, the wire is expected to have a completely gapped spectrum, except for a single Majorana bound state at each end.\cite{Oreg+2010,Lutchyn+2010,Sela+2011} There have been several suggestive experimental signatures of Majorana physics, including zero-bias peaks\cite{Deng+2012,Mourik+2012,Wesperen+2013,Nadj+2014} and an unconventional $4\pi$ Josephson effect,\cite{Wiedenmann+2016} but there is not yet any completely conclusive measurement which would rule out all alternative explanations of the observed features. In this respect, it is necessary to independently characterize the RSOC strength of nanowires. This is one motivation why ``bare'' Rashba nanowires are currently actively studied even in the absence of a superconductor.

Given that proximity to a superconductor screens long-range interactions, experiments on nanowires not subject to the proximity effect should be more sensitive to the interplay between RSOC and electron-electron interactions. A crucial prerequisite for the appearance of Majorana modes is the opening of a partial gap in the spectrum at the Dirac point (see Fig.~\ref{fig:Umklapp}) due to an applied magnetic field. Indeed, the creation of such a gap can be understood most easily for non-interacting electrons, and the fate of this ``helical gap'' in the interacting case has been studied in recent years using renormalization-group (RG) arguments, numerical simulations, and Wigner crystal theory \cite{Braunecker+2010,Fisher+2011,schmidt13b,Schmidt+2016}.

Evidently, an applied magnetic field breaks time-reversal symmetry. However, even without magnetic field when the system is time-reversal symmetric, it is consistent to allow for another spin non-conserving process, so-called spin-umklapp scattering, which can also open a partial gap near the band crossing. While this type of scattering has been predicted based on symmetry arguments, much less attention has been devoted to understanding the microscopic mechanism behind it. Therefore, we will describe in detail how such spin-flipping transitions emerge as a consequence of the quasi-1D nature of experimental Rashba wires. In tandem with electron-electron interactions, they lead naturally to spin-umklapp scattering and hence to the formation of a helical gap even without an applied magnetic field. The resulting system then hosts fractionalized quasiparticles which have attracted a lot of interest in the past years.\cite{kane02,sela11,teo14,oreg14,orth15,ziani15,cavaliere15,cornfeld15}

Therefore, in an interacting quasi-1D Rashba wire, helical gaps can in principle be created by either a magnetic field or by spin-umklapp scattering. This prompts the question about whether the two possible underlying causes can be distinguished experimentally. Both mechanisms lead to very similar experimental signatures in conductance measurements, where the opening of a helical gap results in a halving of the conductance as the chemical potential is tuned close to the Dirac point.\cite{Quay+2010,Zumbuhl+2014} However, we will show that dynamic response functions, in particular the spectral function, the structure factor, and the tunnelling density of states, allow one to uniquely assign one of these origins to an observed helical gap. These functions are thus a much better probe of the helical gap than conductance measurements alone.

The structure of this article is as follows: In Sec.~\ref{sec:results}, we summarize our main results in a non-technical manner. In particular, we define the response functions studied and sketch how helical gaps manifest themselves in them. In Sec.~\ref{sec:model}, we present our model for a quasi-1D Rashba nanowire and show how spin-umklapp scattering is generated by virtual transitions between subbands. In Sec.~\ref{sec:bosonization}, we bosonize the model Hamiltonian to account for the interactions before performing a renormalization group analysis in Sec.~\ref{sec:RG}. Subsequently, in Sec.~\ref{sec:response}, we study the response functions by using the exact form factors of the sine-Gordon model as well as a refermionization solution at certain ``Luther-Emery'' points in parameter space. Finally, we summarize our results and present a short outlook in Sec.~\ref{sec:conclusions}.

\section{Main results}\label{sec:results}

In a strictly 1D nanowire in $x$ direction, RSOC leads to a term $H_R = - \alpha_R \hat{\sigma}_y p_x$ \cite{Gritsev+2005,Lutchyn+2010,Oreg+2010} in the Hamiltonian where $\hat{\sigma}_{x,y,z}$ denotes the Pauli matrices which act on the electron's spin. Adding the kinetic energy $H_0 = p_x^2/(2m)$, one finds that the only effect of RSOC is to shift the parabolic spectra of the two spin components in energy and momentum to $\epsilon_{\uparrow,\downarrow}(p_x) = (p_x \mp m\alpha_R)^2/(2m) - m \alpha_R^2/2$. Importantly, however, the 1D Hamiltonian still commutes with the $y$-component of the electron's spin, $[H_0 + H_R, \hat{\sigma}_y] = 0$. Hence, merely adding Coulomb interactions to a 1D Rashba Hamiltonian is not sufficient to create spin-flip processes, since in that case the total spin of the electrons will still be conserved.

However, realistic nanowires have a physical extent in a second direction, making them quasi-1D structures. They can be modelled by narrowly confining a two-dimensional system with a harmonic trapping potential. This results in transverse subbands separated in energy by the trapping frequency $\Omega$. For a 2D system in the $x-y$ plane, the Rashba Hamiltonian reads $H_R = \alpha_R ( \hat{\sigma}_x p_y - \hat{\sigma}_y p_x)$ and, in contrast to the 1D version, no longer commutes with $\hat{\sigma}_y$. From this form of RSOC, one finds that transitions between neighbouring subbands are always associated with a spin flip. By analogy to an applied magnetic field perpendicular to the Rashba spin-orbit axis, it has also been discussed how inter subband transitions can induce gaps in the spectrum.\cite{Meng+2014}

\begin{figure}[t]
	\centering
	\includegraphics[width=\columnwidth]{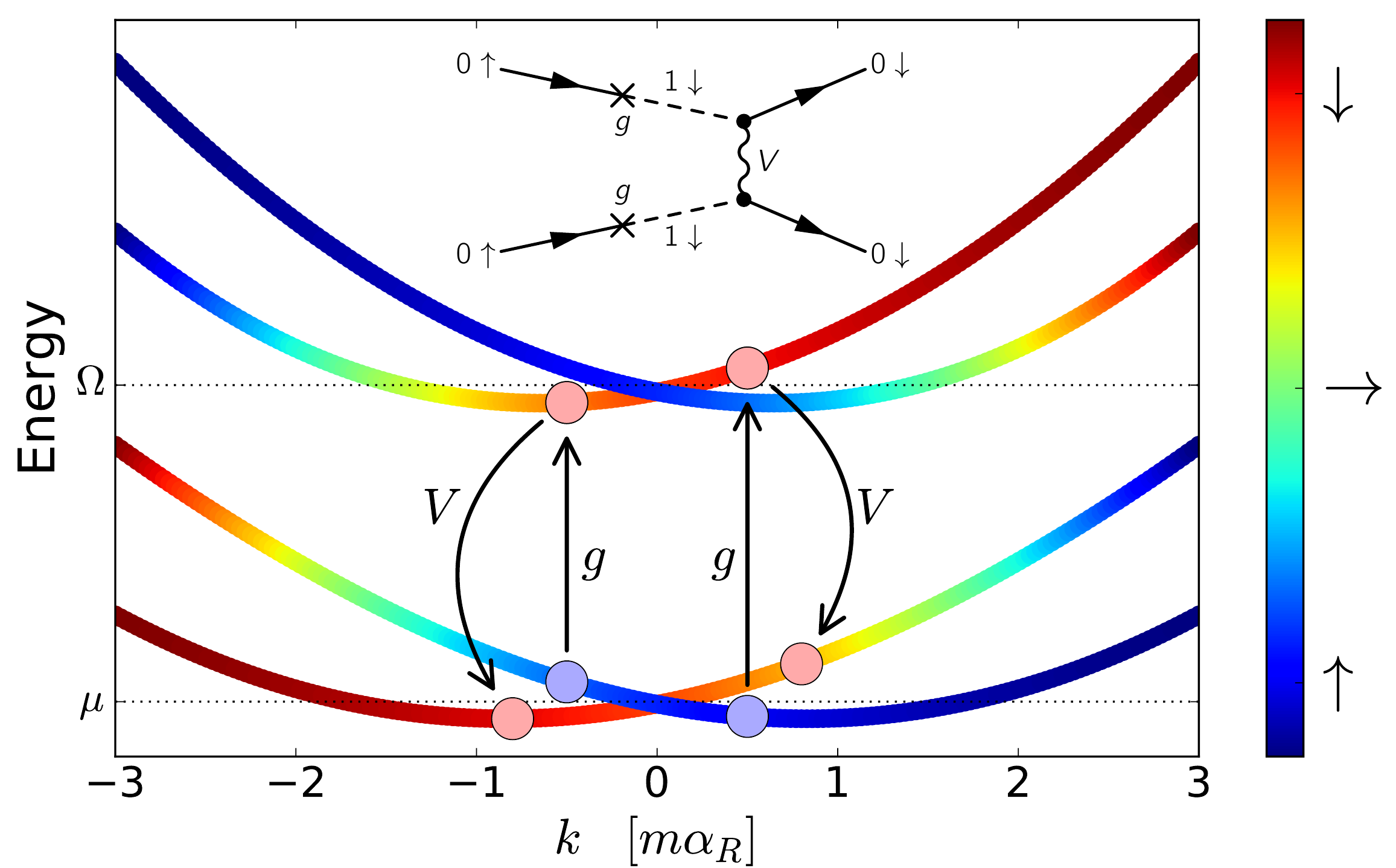}
	\caption{Single-particle spectrum of a Rashba wire with two transverse channels in the absence of magnetic field. The spin texture is indicated by the color coding. Virtual transitions between subbands conserve momentum and are associated with a spin flip. In combination with electron-electron interaction, this can produce spin-umklapp scattering as indicated by the Feynman diagram on top.}
	\label{fig:Umklapp}
\end{figure}

The effect of these subbands on transport in a wire with RSOC and electron-electron interactions has previously been studied in a series of papers.\cite{Moroz+2000a,Moroz+2000b} However, these studies focussed on the gapless modes present in the system, and neglected the effect of magnetic fields or backscattering on transport. When combined with strong enough Coulomb interactions, we find that virtual transitions between subbands generate a spin-umklapp scattering process in which, e.g., two spin-up electrons are converted into two spin-down electrons. Hence, this process changes the total spin without breaking time-reversal symmetry. This process is schematically depicted in Fig.~\ref{fig:Umklapp}.

To fully analyse this spin-umklapp scattering and study its interplay with a magnetic field, we carry out a first-order renormalization group (RG) analysis. We find that whereas a magnetic field perpendicular to the Rashba axis always opens a partial gap in the spectrum, it is only for sufficiently strong interactions, characterized by a Luttinger parameter $K<1/2$, that spin-umklapp scattering can open a similar gap, provided the chemical potential is tuned close to the Dirac point. However, such strong interactions ($K \sim 0.2-0.4$) have been observed in semiconductor nanowires \cite{Jompol+2009,Hevroni+2015} and carbon nanotubes.\cite{McEuen+1999}

As both a perpendicular magnetic field and spin-umklapp scattering open a partial gap, we must conceive of a means of telling them apart. As we will show below, one crucial difference is that with a magnetic field-induced gap, the low-energy excitations near to $k=0$ (``quasiparticles") of the system have the same quantum numbers as the original electrons, whereas in the case of spin-umklapp scattering they have fractionalized charge $e/2$. This makes both effects distinguishable in response functions. 

For instance, exciting quasiparticles across the band gap requires an energy $\omega>\Delta$ regardless of how the gap $\Delta$ is generated. This corresponds to the gap which can be measured for instance in the density structure factor,
\begin{align}\label{eq:DSF}
	S (k,\omega) &= \int dx dt e^{i\omega t-ikx} \left\langle \rho (x,t) \rho (0,0) \right\rangle ,
\end{align}
where $ \rho = \sum_\sigma \, \rho_\sigma$ is the total electron density, and $\sigma = \uparrow, \downarrow$.

By contrast, inserting a single physical electron into the system will require producing a \emph{single} charge-$e$ quasiparticle if the gap is generated by a magnetic field, but \emph{two} charge-$e/2$ excitations if it is caused by spin-umklapp scattering. Experimentally, this can be probed in the spectral function
\begin{equation}\label{eq:SF}
    A_{\psi}(k,\omega) = \frac{1}{2\pi} \int dx dt e^{i\omega t-ikx} \left\langle \left\{ \psi (x,t),  \psi^\dag (0,0)\right\} \right\rangle,
\end{equation}
where $\psi = \cos(\theta/2) \psi_\uparrow + e^{i \phi} \sin(\theta/2) \psi_\downarrow$ is the annihilation operator for an electron with spin pointing in a direction on the Bloch sphere given by the azimuthal angle $\phi \in [0, 2\pi]$ and the polar angle $\theta \in [0, \pi]$. Moreover, $\{\cdot,\cdot\}$ denotes the anticommutator. The gap seen in $A_\psi(k,\omega)$ would also be visible in the d.c. conductance of the wire, and the tunneling density of states,\cite{Braunecker+2012,Schuricht+2012}
\begin{align}
    \rho_\psi(\omega) = \int dk A_\psi(k,\omega) ,
\end{align}
which can be measured experimentally through spin-polarized scanning tunneling microscopy.

\begin{figure}[t]
	\centering
	\includegraphics[width=\columnwidth]{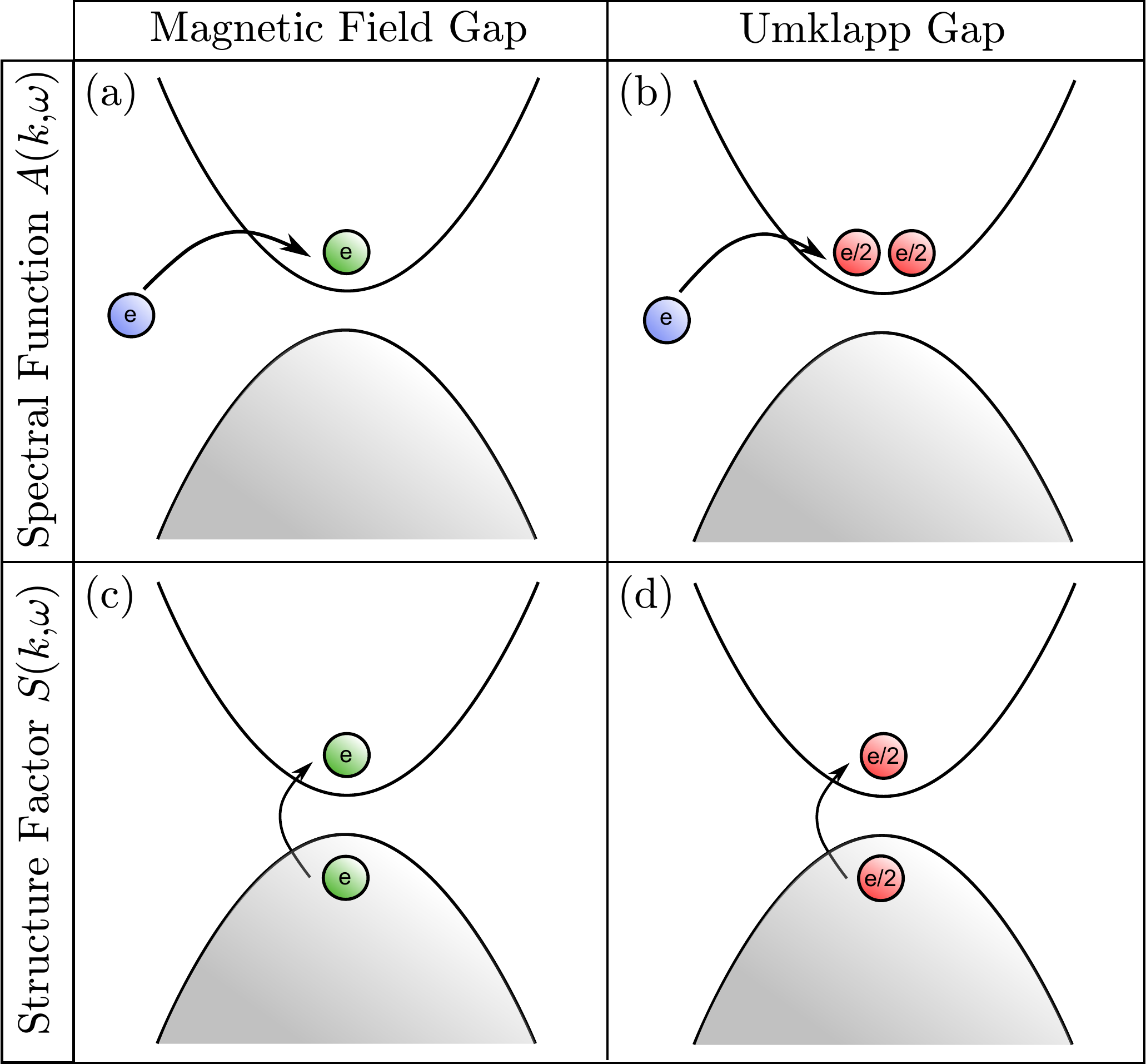}
	\caption{A schematic representation of the various response functions we consider. Panel (a) shows the insertion of a single electron into a system with a $B$-field gap, resulting in a single quasiparticle of mass $M_1$ in the ``conduction'' band. This will result in threshold behavior for the spectral function at $\omega=M_1$. Panel (b) shows the same process, but in the case of an umklapp gap, where in order to obey charge conservation we must create \emph{two} $e/2$ quasiparticles of mass $M_2$, so the threshold value shifts to $\omega=2M_2$. Panels (c) and (d) show the process interrogated by the density structure factor, which will have the same threshold behavior, regardless of the underlying gap-generating mechanism.}
	\label{fig:response_functions}
\end{figure}

By calculating the spectral function and the density structure factor for the gapped modes at $k=0$ and the gapless modes at $k=\pm k_F$, we find that the simple picture we just presented is borne out. In particular, the spectral function onsets at different values of energy, and also takes a different functional form depending on the gap-opening mechanism. These results are summarized in Fig.~\ref{fig:response_functions}.

Even though we provide a specific microscopic derivation of the spin-umklapp scattering in quasi-1D wires, we would like to stress that the mechanism we describe for opening of a helical gap without a magnetic field is rather generic, and relies only on sufficiently strong interactions and clean wires. Therefore, our predictions about the response functions remain valid even if the spin-umklapp term is generated by other mechanisms than the inter-subband transitions explained below.

\section{Model}\label{sec:model}

We consider a model of the experimental arrangement used to investigate the conductance of Rashba spin-orbit coupled quantum wires. Our prototypical system has a large extent in the $x$-direction, but is confined in the $y$-direction by a harmonic potential of strength $\Omega$ to model the effect of the finite width of the real system. The dominant spin-orbit coupling is of Rashba type, resulting from the breaking of structural inversion symmetry associated to either an intrinsic or an externally-applied electric field $\vec{E} \parallel \hat{z}$. The RSOC then has strength $\alpha_R$ proportional to $|\vec{E}|$. The Hamiltonian for the system takes the form \cite{Moroz+2000a,Moroz+2000b,Starykh+2008,Pedder+2015}
\begin{equation}
H = \frac{p_x^2 + p_y^2}{2m} + \frac{1}{2} m \Omega^2 y^2 + \alpha_R (\hat{\sigma}_x p_y - \hat{\sigma}_y p_x) ,
\label{RashbaHamiltonian}
\end{equation}
where $p_x$, $p_y$ are the momenta parallel and perpendicular to the wire axis, and $\hat{\sigma}_x$ and $\hat{\sigma}_y$ are two spin-$1/2$ Pauli matrices. The harmonic confinement along the $y$-direction suggests introducing raising and lowering operators $a^\dagger$ and $a$, where $y = (2 m \Omega)^{-1/2}(a + a^\dagger)$. The Hamiltonian then decomposes into two parts $H = H_0 +H_1$, where
\begin{align}
H_0 &= \Omega \left( a^\dagger a + \frac{1}{2}\right) + \frac{p_x^2}{2m} - \alpha_R \hat{\sigma}_y p_x, \label{H0} \\
H_1 &= i g \hat{\sigma}_x (a^\dagger - a), \label{H1}
\end{align}
where we have defined $g = \alpha_R \sqrt{m \Omega/2}$. Hence, the motion in $y$-direction results in a tower of subbands which come from the finite width of our model. Since $\hat{\sigma}_y$ commutes with $H_0$, the eigenstates of $H_0$ have definite spin in the $y$-direction. It is then clear that transitions between subbands, created by the raising and lowering operators in $H_1$, flip this spin because $H_1$ contains $\hat{\sigma}_x$.

We are mainly interested in the low-energy behavior of this system, so we focus only on the lowest and the first-excited subbands. In order to generate an effective model which accounts for the physics arising from changes of subband, we perform a Schrieffer-Wolff transformation to remove the coupling $H_1$ between the subbands up to order $(g/\Omega)^2$. We conjugate the Hamiltonian $H$ by a transformation $H^\prime = e^{-S} H e^S$, where we choose
\begin{equation}
S=-\frac{g (a^\dag \hat{\sigma}^- - a \hat{\sigma}^+)}{(2\Omega - 4 \alpha_R p_y)} + \frac{g (a^\dag \hat{\sigma}^+ - a \hat{\sigma}^-)}{(2\Omega + 4 \alpha_R p_y)} ,
\end{equation}
such that $[S,H_0]=-H_1$. We can see from this expression that $S$ is $\mathcal{O}(g)$, and expanding the transformation to $\mathcal{O}(g^2)$, we find
\begin{align}
H^\prime &= H_0  - \frac{1}{2}[S,H_1] + \mathcal{O}(g^3) \notag \\
& \approx \frac{p_x^2}{2m} - \alpha_R \hat{\sigma}_y p_x + \epsilon_{\text{SO}} ,
\end{align}
where we discard the constant shift $\epsilon_{\text{SO}}$ of the total energy. Ultimately this analysis tells us that, to this order, it would have been consistent to ignore the dynamics in the $y$-direction in Eq.~(\ref{RashbaHamiltonian}) from the beginning, and simply use the transformed Hamiltonian $H^\prime$, as we would expect for non-interacting electrons. Note that deviations of the spectrum from the parabolic form occur at higher orders in $(g/\Omega)$,\cite{Moroz+2000a,Moroz+2000b,Kainaris+2015} but they are not important for our analysis.

The typical screened Coulomb interactions between electrons create a generic density-density interaction potential $V(\vec{r}_1 - \vec{r}_2)$. This interaction conserves the two spins of both interacting electrons. In second-quantized language, the interaction term takes the form
\begin{align}
\hat{V} = \sum_{\sigma_1\sigma_2} \int d\vec{r}_1 d\vec{r}_2   \psi_{\sigma_1}^\dagger (\vec{r}_1) \psi_{\sigma_2}^\dagger (\vec{r}_2) V (|\vec{r}_1-\vec{r}_2|)\psi_{\sigma_2}(\vec{r}_2) \psi_{\sigma_1}(\vec{r}_1),
\label{PotentialTerms}
\end{align}
where the operator $\psi_{\sigma}(\vec{r})$ annihilates an electron with spin $\sigma = \uparrow,\downarrow$ (in $y$-direction) at position $\vec{r}$. Transforming the interaction term $V$ using the Schrieffer-Wolff approach described above, and projecting into the lowest subband, we find that the second-quantized Hamiltonian in momentum space takes the form $\hat{H} = \hat{H}_0 + \hat{V}_\rho + \hat{V}_{\text{sf}} + \hat{V}_{\text{sx}}$, where $\hat{H}_0$ is
\begin{equation} \label{eq:hamzero}
\hat{H}_{0} = \sum_{\sigma,p} \left( \frac{p^2}{2m} - \alpha_{\rm{R}} \sigma p \right) \psi^\dagger_{p,\sigma} \psi_{p,\sigma}.
\end{equation}
The operators $\psi_{p,\sigma}^\dagger $ and $\psi_{p,\sigma}$ create and annihilate, respectively, an electron with spin $\sigma$ and momentum $p$ (in $x$ direction), which is a good quantum number as a result of translational invariance in the $x$-direction.

\begin{figure}[t]
	\centering
	\includegraphics[width=\columnwidth]{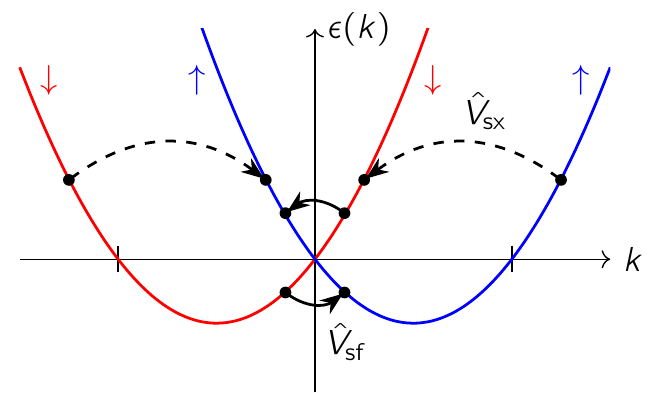}
	\caption{Effective interactions in the lowest subband due to virtual transitions to higher subbands: spin-flip terms $\hat{V}_{\rm sf}$ (denoted by solid arrows) and spin-exchange terms $\hat{V}_{\rm sx}$ (denoted by dashed arrows).}
	\label{fig:InteractionProcesses}
\end{figure}

The terms arising from the transformation in Eq.~(\ref{PotentialTerms}) are a density-density interaction $\hat{V}_\rho$, a spin-flip term $\hat{V}_{\text{sf}}$, and a spin exchange term $\hat{V}_{\text{sx}}$. The interaction $\hat{V}_\rho$ is given by
\begin{equation} \label{eq:Vrho}
\hat{V}_\rho = \frac{1}{L} \sum_{\sigma_1 \sigma_2} \sum_{p,p^\prime,q} \tilde{V} (q) \psi^\dagger_{p+q,\sigma_1} \psi^\dagger_{p^\prime-q,\sigma_2} \psi_{p^\prime,\sigma_2} \psi_{p,\sigma_1},
\end{equation}	
where $L$ is the length of the wire. Moreover, $\tilde{V} (q) = \tilde{V}(q,y=0)$, where $\tilde{V}(q,y)$ is a partial Fourier transformation of the interaction potential $V(|\bf{r}|)$: $\tilde{V}(q,y) = \int dx e^{-i q x} V(x, y)$. The remaining two terms may be written
\begin{align}
&\hat{V}_{\rm{sf}/\rm{sx}}  \label{eq:Vsfsx}\\
&= \frac{1}{L} \underset{\sigma}{\sum} \underset{p,p^\prime,q}{\sum} \tilde{U} (q) (2p^\prime - q)(2p+q) \psi^\dagger_{p+q,\sigma} \psi^\dagger_{p^\prime-q,\pm \sigma} \psi_{p^\prime,\mp \sigma} \psi_{p,-\sigma}, \nonumber
\end{align}
with the effective interaction potential
\begin{align}
\tilde{U}(q)
&=	\frac{m \alpha_R^4}{2 \pi^{3/2}\Omega^3}
\overset{\infty}{\underset{n=0}{\sum}} \frac{1}{\sqrt{2^n n!}} \int_{-\infty}^\infty dz_1 dz_2 dz_3 e^{-z_1^2-z_2^2-\frac{(z_1-z_2)^2}{2}} \nonumber\\
& \times  H_{n}(z_1 - z_2) H_1(z_1) H_1(z_2) e^{-z_3^2/2} H_n(z_3) \tilde{V} \left(q, \frac{z_3}{\sqrt{m \Omega}} \right).
\label{uofq}
\end{align}
The functions $H_n (z)$, which are $n$th order Hermite polynomials, occur as a result of the harmonic confinement in the $y$-direction. The spin-exchange term $\hat{V}_{\rm{sx}}$ describes scattering within the lowest subband in which the interacting electrons exchange their spin. In contrast, the spin-flip term $\hat{V}_{\rm{sf}}$ corresponds to a process in which both of the interacting particles flip their spin (see Fig.~\ref{fig:InteractionProcesses}). Hence, this latter term contains spin-umklapp scattering which changes the total spin by two.

To deal with the case of Rashba wires in an applied magnetic field, we include a Zeeman term which couples to the spin of the electrons,
\begin{align}
\hat{H}_{\text{B}}
&= \sum_{i=y,z} \underset{p,\sigma,\sigma^\prime}{\sum} B_i  \psi^\dagger_{p,\sigma} \hat{\sigma}_{\sigma \sigma^\prime}^i \, \psi_{p,\sigma^\prime} \notag \\
&= \underset{p}{\sum} \left[ \underset{\sigma}{\sum}  B_y \sigma \psi^\dagger_{p,\sigma}\psi_{p,\sigma} + B_z \left( \psi_{p,\uparrow}^\dag \psi_{p,\downarrow} + \text{h.c.}  \right) \right].
\end{align}
where $\hat{\sigma}^i$ is a triplet of Pauli matrices. Since the spin is quantized along the $y$-direction, we chose $\hat{\sigma}^y = \text{diag}(1, -1)$ to be diagonal, and $\hat{\sigma}^{z,x}$ then follow from the spin commutation relations. As we restrict ourselves to states in the lowest subbands, a general magnetic field can be split into two important components, parallel and perpendicular to the Rashba axis taken, without loss of generality, as along the $y$-direction, and the $z$-direction respectively. Whilst it is well-known that the component of the magnetic field perpendicular to the Rashba axis is the one which opens the partial gap in Majorana nanowires, we keep also the parallel component in order to check to what extent small stray fields along this direction can destabilize the helical gap.

Moreover, a back-gate present in most experiments gives control over the population of electrons in the wire. It can be used to sweep the chemical potential, and so can be experimentally tuned to specific interesting values (e.g.~the Dirac point of the system). The chemical potential couples to the density, and so appears in the Hamiltonian according to
\begin{align}
\hat{H}_{\mu}
&= - \mu \sum_{p,\sigma}  \psi^\dagger_{p,\sigma}\psi_{p,\sigma} .
\end{align}
Hence, the complete Hamiltonian describing the electrons in the lowest subband and accounting to leading order for virtual transitions to the first excited subband is given by
\begin{align}\label{Hfull}
\hat{H} = \hat{H}_0 + \hat{H}_B + \hat{H}_\mu + \hat{V}_\rho + \hat{V}_{\rm sf} + \hat{V}_{\rm sx} .
\end{align}
We proceed by bosonizing this Hamiltonian and using the bosonized version as basis for a renormalization group analysis.

\section{Bosonization}\label{sec:bosonization}

To study the effect of the various interaction terms, we analyse our system using bosonization. We restrict our attention to small energies near to the chemical potential $\mu$, and assume that the latter is tuned to be close to the Dirac point (see Fig.~\ref{fig:InteractionProcesses}). The kinetic energy Eq.~(\ref{eq:hamzero}) gives rise to two parabolic spectra for spin-$\uparrow$ and spin-$\downarrow$ electrons, which are shifted relative to each other in momentum space by RSOC. Placing the chemical potential at the Dirac point, we have a total of four low-energy modes: two with opposite spins at $k=0$ and two with opposite spins at $k = \pm k_F$, where $k_F=2m\alpha_R$. To study the low-energy sector, we therefore project the field operators $\psi_\sigma(x)$ as follows,
\begin{align}
\psi_\uparrow (x) &\approx e^{ik_F x} \psi_{R \uparrow} (x) + \psi_{L \uparrow} (x), \nonumber \\
\psi_\downarrow (x) &\approx \psi_{R \downarrow} (x) + e^{-ik_F x} \psi_{L \downarrow} (x) . \label{eq:lowemodes}
\end{align}
We define the Fourier components of these fields as
\begin{equation}
\psi_{\alpha \sigma}(x) = \frac{1}{\sqrt{L}} \underset{k}{\sum} e^{ikx} \psi_{\alpha \sigma,k},
\end{equation}
where $\alpha = R,L$ and $\sigma=\uparrow,\downarrow$, and the summation over $k$ is taken over a narrow momentum window $|k| < \Lambda$, where $\Lambda \ll k_F$ is the momentum cutoff.

Next, we turn our attention to projecting the different terms in the full Hamiltonian Eq.~(\ref{Hfull}) onto these low-energy degrees of freedom. Starting with the spin-flip terms in Eq.~(\ref{eq:Vsfsx}) we find
\begin{align}
\hat{V}_{\text{sf}} = \frac{\tilde{U}(0)}{L} \underset{pp^\prime q}{\sum} & (2p^\prime-q)(2p+q) \biggl[\psi_{L \uparrow,p+q}^\dag \psi_{L \uparrow,p^\prime-q}^\dag \psi_{R\downarrow,p} \psi_{R\downarrow,p^\prime} \nonumber \\
& + \psi_{R \downarrow,p+q}^\dag \psi_{R \downarrow,p^\prime-q}^\dag \psi_{L\uparrow,p} \psi_{L\uparrow,p^\prime} \biggr] , \label{eq:Vspinflip}
\end{align}
where we have used the small range of the momenta to justify neglecting the momentum dependence of the potential $\tilde{U}(q)$. The interpretation of such a term is that it converts two spin-$\uparrow$ left-movers into two spin-$\downarrow$ right-movers, or vice versa, thus only conserves spin modulo 2. Note that an analogous process which would scatter two spin-$\downarrow$ left-movers into two spin-$\uparrow$ right-movers is not present because it would change momentum by $4 k_F$, which is not allowed in our translation-invariant system.

\begin{figure}[t]
	\centering
	\includegraphics[width=\linewidth]{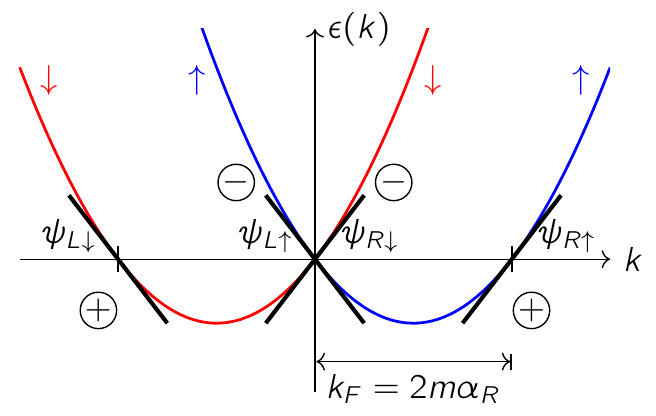}
	\caption{This figure shows the position of the chemical potential and the four low-energy linearized modes used to bosonize the system. When bosonized, the inner modes near $k = 0$ are described by the operators ($\phi_-,\theta_-$), the outer modes near $k = \pm k_F$ by ($\phi_+,\theta_+)$.}
	\label{fig:Linearized}
\end{figure}

We continue with the spin-exchange terms $\hat{V}_{\text{sx}}$ in Eq.~(\ref{eq:Vsfsx}). While those mostly yield merely corrections to the terms already present in the density-density interaction term $\hat{V}_\rho$, there remains one term which cannot be written as a density-density interaction and must thus be kept separate, namely
\begin{align} \label{eq:Vcooper}
\hat{V}_{\text{S}} = -\frac{2 k_F^2 \tilde{U}(k_F)}{L} \underset{p,p^\prime,q}{\sum} & \biggl[ \psi_{L\uparrow,p+q}^\dag \psi_{R\downarrow,p^\prime-q}^\dag \psi_{R\uparrow,p^\prime} \psi_{L\downarrow,p}  \\
& + \psi_{L\downarrow,p+q}^\dag \psi_{R\uparrow,p^\prime-q}^\dag \psi_{R\downarrow,p^\prime} \psi_{L\uparrow,p} \biggr] . \nonumber
\end{align}
This process couples electrons near $\pm k_F$ with electrons near the Dirac point, and exchanges their spins. It is graphically depicted in Fig.~\ref{fig:InteractionProcesses}.

Moreover, we also write the magnetic field and the chemical potential Hamiltonians in terms of the low-energy degrees of freedom
\begin{align}
\hat{H}_B &=
\int \, dx \, \biggl\{ B_y [\rho_\uparrow (x) - \rho_\downarrow (x)]
+ B_z \left[ \psi_{L \uparrow}^\dag (x) \psi_{R \downarrow} (x) + \text{h.c.} \right] \biggr\},\notag\\
\hat{H}_\mu &=
-\mu \int \, dx \left[\rho_\uparrow (x) + \rho_\downarrow (x)\right] ,
\end{align}
where we have defined the spin-resolved electron densities $\rho_{\sigma} = \rho_{R \sigma} + \rho_{L \sigma}$ ($\sigma = \uparrow,\downarrow$) in terms of the density operators $\rho_{\alpha \sigma}(x) = \psi_{\alpha \sigma}^\dag (x) \psi_{\alpha \sigma}(x)$. We now bosonize the complete Hamiltonian using the bosonization identities
\begin{equation} \label{eq:bosons}
\psi_{\alpha \sigma} (x) = \frac{U_{\alpha \sigma}}{\sqrt{2 \pi a}} e^{-i\varphi_{\alpha \sigma}(x)},
\end{equation}
where the chiral bosonic fields are defined as $\varphi_{\alpha \sigma} = \alpha \phi_{(\alpha \sigma)} - \theta_{(\alpha \sigma)}$ for $\alpha = R,L = +,-$ and $\sigma = \uparrow,\downarrow = +,-$. Here, the composite index $(\alpha \sigma) = \pm$ denotes the product of $\alpha$ and $\sigma$, such that, e.g., $\varphi_{L\uparrow} = - \phi_{-} - \theta_-$ and $\varphi_{L\downarrow} = -\phi_+ - \theta_+$. This notation ensures that the canonically conjugate pair of fields $(\phi_-,\theta_-$) labels the modes near $k =0$, whereas $(\phi_+, \theta_+)$ labels the modes near $k = \pm k_F$ (see Fig.~\ref{fig:Linearized}).

The operators $U_{\alpha \sigma}$ are the Klein factors needed to ensure that fermions on different branches obey the correct anticommutation relations, and $a \approx 1/\Lambda$ is a short-range cut-off. The Klein factors play no further role in the physics, so we will drop them from here on. With the definitions (\ref{eq:bosons}), the density operators become
\begin{align} \label{eq:densities}
\rho_{\alpha \sigma} (x) &= -\frac{\alpha}{2 \pi} \partial_x \varphi_{\alpha \sigma}(x) .
\end{align}
Rewriting the kinetic energy term (\ref{eq:hamzero}) in terms of these bosonic variables, we get two uncoupled Luttinger liquid Hamiltonians for the $(\phi_+,\theta_+)$ and $(\phi_-,\theta_-)$ modes. Including the density-density terms arising from Eqs.~(\ref{eq:Vrho}) and (\ref{eq:Vsfsx}), we find that there are also derivative terms which couple the two species, and renormalize the sound velocities and the Luttinger parameters for the two different $\pm$ bosonic species.

The derivative terms coupling the two species can be removed by going to the charge-spin basis of Ref.~[\onlinecite{Starykh+2008}] by defining $\phi_{\rho,\sigma} = (\phi_+ \pm \phi_-)/\sqrt{2}$ and $\theta_{\rho,\sigma} = (\theta_+ \pm \theta_-)/\sqrt{2}$. The kinetic part of the Hamiltonian now reads
\begin{equation}
\hat{H}_0 = \underset{j=\rho,\sigma}{\sum} \frac{v_j}{2 \pi} \int \, dx \left( \frac{(\partial_x \phi_j)^2}{K_j} + K_j (\partial_x \theta_j)^2 \right),
\end{equation}
where $v_j$ and $K_j$ are the sound velocities and Luttinger parameters of the charge and spin modes, respectively. For repulsive interactions one finds $K_\rho <1$ and $K_\sigma \leq 1$ ~[\onlinecite{Starykh+2008}]. In the case of unbroken SU$(2)$ spin-rotation symmetry, we would be restricted to $K_\sigma =1$, however the Rashba spin-orbit coupling in combination with an applied magnetic field breaks this symmetry, leading to $K_\sigma<1$. Note that this is in contrast to the usual case of degenerate branches of spinful fermions without RSOC, where for repulsive interactions $K_\sigma >1$.\cite{Giamarchi+2004}

The two competing terms arising from the interaction terms $\hat{V}_{\rm S}$ and $\hat{V}_{\rm sf}$ in bosonic language become
\begin{align} \label{eq:Vsxboson}
\hat{V}_{\text{S}} &= g_{\text{S}} \int \, dx \cos [2 \sqrt{2} \theta_\sigma] , \\
\label{Vsfboson} \hat{V}_{\text{sf}} &= g_{\text{U}} \int \, dx \cos [2 \sqrt{2}(\phi_\rho - \phi_\sigma)].
\end{align}
For weak interactions, the coupling constants $g_{\text{S}}$ and $g_{\text{U}}$ are given by $g_{\text{S}} =2k_F^2 \tilde{U}(k_F)/(2\pi)^2$ and $g_{\text{U}} = -2\tilde{U}(0)/(2\pi a)^2$, but for stronger interactions it will be more convenient to treat them as phenomenological parameters.

Finally, the Hamiltonians due to the magnetic field and the chemical potential become
\begin{align} \label{eq:hextboson}
\hat{H}_B &= -\frac{\sqrt{2} B_y}{\pi} \int \, dx \, \partial_x \theta_\sigma + \frac{B_z}{\pi a} \int \, dx \, \cos (\sqrt{2}[\phi_\rho - \phi_\sigma]) , \notag \\
\hat{H}_\mu &= \frac{\sqrt{2} \mu}{\pi} \int \, dx \, \partial_x \phi_\rho .
\end{align}

\section{Renormalization group analysis}\label{sec:RG}

To find out which terms dominate the behavior of the system and to assess the possibility of opening gaps due to spin-umklapp scattering and a magnetic field, we investigate the flow of the three cosine terms, which have coupling constants $g_{\text{S}}$, $g_{\text{U}}$ and $B_z$. We carry out our analysis first precisely at the Dirac point of the system, $\mu = \mu^\star$, where we expect umklapp scattering to be potentially relevant, before considering small shifts of the chemical potential $\mu$ and parallel magnetic field $B_y$ away from this special point. We study the flow of these terms, and investigate their potential to spoil the gaps generated by the cosine terms.

We use a real space renormalization group (RG) analysis based on the operator-product expansion.\cite{Cardy+1996} A similar analysis has been carried out by Stoundenmire \emph{et al.}, \cite{Fisher+2011} However, they focussed their attention on the effect of generic two particle backscattering on the appearance of a topological superconducting phase, which relies on the whole system being in a gapped phase. In contradistinction, we concentrate on the opening of a helical gap either by magnetic field or umklapp scattering, but without proximity to an $s$-wave superconductor.

In order to compute the first-order RG equations, we calculate the scaling dimensions of each of the interaction terms in the Hamiltonian separately. We find for the cosine terms that
\begin{align}
\frac{d g_{\text{S}}}{d \ell} &= \left(2-\frac{2}{K_\sigma} \right) g_{\text{S}} , \\
\frac{d g_{\text{U}}}{d \ell} &= 2\left(1-K_\sigma -K_\rho \right) g_{\text{U}} , \\
\frac{d B_z}{d \ell} &= \left(2-\frac{K_\sigma +K_\rho}{2} \right) B_z ,
\end{align}
where the flowing cutoff is parameterized as $a=e^{-\ell} a_0$ where $\ell$ flows from $0$ to $\infty$ as we go to lower energies. These RG equations demonstrate that the spin-flip term is irrelevant for repulsive interactions where $K_\sigma < 1$. In contrast, the umklapp term $g_{\rm U}$ can become relevant for strongly repulsive interactions, where $K_\rho + K_\sigma < 1$, in which case it will flow to strong coupling and open a partial gap. The perpendicular magnetic field $B_z$ flows to strong coupling for $K_\rho + K_\sigma < 4$. This means that for all repulsive interactions, $B_z$ is more strongly relevant than the umklapp term.

For the terms corresponding to a magnetic field and a chemical potential, we find
\begin{align}
\frac{d \mu}{d \ell} &= \mu , \\
\frac{d B_y}{d \ell} &= B_y .
\end{align}
The chemical potential term $\mu$ and the field along the Rashba direction $B_y$, in contrast to the terms found before, flow trivially with the cutoff, and so are also RG-relevant, albeit less so than the $B_z$-field term.

Despite this trivial scaling, these terms can still have physical significance for the opening of the gap, and in principle are capable of rendering other \emph{a priori} RG-relevant terms discovered above irrelevant in the thermodynamic limit. We investigate this by allowing for a small shift of the chemical potential away from the Dirac point.

Going back to the bosonized expression in Eq.~(\ref{eq:hextboson}), it is clear that we may remove the linear derivative terms corresponding to a chemical potential shifted away from the Dirac point, and a weak parallel magnetic field by shifting the fields $\phi_\rho$ and $\theta_\sigma$ by an $x$-dependent quantity according to
\begin{align} \label{eq:deltarho}
\tilde{\phi}_\rho (x) &= \phi_\rho(x) + \frac{\sqrt{2} \mu}{v_\rho K_\rho} x = \phi_\rho(x) + \delta_\rho x , \\
\label{eq:deltatheta} \tilde{\theta}_\sigma (x) &= \theta_\sigma (x) - \frac{\sqrt{2} B_y K_\sigma}{v_\sigma} x = \theta_\sigma(x) - \delta_\sigma x .
\end{align}
The linear derivative terms are absorbed into the quadratic terms, at the expense of throwing away a constant contribution to the kinetic part of the Hamiltonian, which becomes
\begin{align}
\hat{H}_0 &= \frac{v_\rho}{2\pi} \int \, dx \left( \frac{1}{K_\rho}(\partial_x \tilde{\phi}_\rho)^2 + K_\rho (\partial_x \theta_\rho)^2 \right) \nonumber \\
& + \frac{v_\sigma}{2\pi} \int \, dx \left( \frac{1}{K_\sigma}(\partial_x \phi_\sigma)^2 + K_\sigma (\partial_x \tilde{\theta}_\sigma)^2 \right).
\end{align}
The linear shift of the term in $\theta_\sigma$ does not appear in any potentially RG-relevant term in the Hamiltonian, and can therefore be ignored. In the umklapp term, the shift of $\phi_\rho$ means the argument of the cosine now acquires a linear dependence on position $\cos [2\sqrt{2} (\phi_\rho(x) - \phi_\sigma(x))] \rightarrow \cos [2\sqrt{2}(\tilde{\phi}_\rho(x) - \phi_\sigma(x)+\delta_\sigma x)]$, which then causes this term to oscillate in space. In the thermodynamic limit, such oscillating terms average to zero, and become irrelevant in the RG sense. This reasoning suggests that any component of the magnetic field along the Rashba axis will prevent an umklapp gap from opening.

We should be more careful, however. Our previous analysis shows that at $\mu=\mu^\star$, the RG relevant cosine term flows to strong coupling, and opens a gap of size $\Delta$. If we now allow $\mu$ to deviate a little from this value $\mu = \mu^\star + \delta \mu$, so long as $\delta \mu < \Delta$, we remain in the gapped phase, and we must consider the effect of the $\delta \mu$ term and the ${B_z}$ and umklapp terms on the same footing. It is clear from the RG equations that, since $K_\rho <1$ and $K_\sigma <1$, $B_z$ will flow to strong coupling ahead of $\mu$ for similar starting values of the two parameters.

Indeed, pinning $\sqrt{2}(\tilde{\phi}_\rho(x) - \phi_\sigma(x)+\delta_\rho x)$ such that the cosine in Eq.~(\ref{eq:hextboson}) is at its minimum value corresponds to adjusting the density $\partial_x \phi_\rho$ by a constant amount, which is acceptable when working at fixed chemical potential, as opposed to at fixed particle number. This adjustment in density does not move us out of the gapped phase, however, and the magnetic field-generated gap survives.

In the case of the umklapp gap, $\mu$ will flow to strong coupling faster than $g_\text{U}$. Since our RG is perturbative, we must cut the flow when one of the coupling constants becomes of order one. The question then becomes one of bare values of the coupling constants, we want the bare value of $g_\text{U}$ to be sufficiently large that it flows to $\mathcal{O}(1)$ before $\mu$. With back-gate control over the chemical potential, tuning to this regime is in principle possible.

Since we are interested in a situation where one of the coupling constants always runs to strong coupling, we might worry about the validity of using an analysis based on perturbative RG. The study of the commensurate-incommensurate (C-IC) transition in terms of classical, and later semi-classical solitons carried out in Refs.~[\onlinecite{Frank+1949,Schulz+1983,Tsvelik+2001}] (for a review, see Ref.~[\onlinecite{Bak+1982}]) shows that our RG-based intuition is correct, and that the gap generated by a relevant cosine term is robust to small off-axis magnetic fields, and small shifts of the chemical potential away from the Dirac point for a finite-sized system.

The main result of this section therefore is that even in the presence of a finite shift of the chemical potential away from the Dirac point, a sufficiently strong magnetic field perpendicular to the Rashba axis opens a partial gap in the spectrum. However, in the absence of a perpendicular magnetic field, and for stronger interactions so that $K_\rho+K_\sigma<1$, umklapp scattering can open a similar gap at the band crossing under similar conditions.

\section{Response Functions}\label{sec:response}

We now turn to the question of how to experimentally differentiate between a partial gap generated by a magnetic field, and one caused by umklapp scattering. Whilst we have many physical probes, e.g., the d.c.~conductance which will see a gap in the spectrum created by the two possible mechanisms, the issue of finding the underlying cause is more subtle.

In particular, one can measure a.c.~conductance or optical properties of the system to get access to the density-density correlations encoded in the dynamical structure factor $S(k,\omega)$ of the wire. These correlations include contributions from exciting quasiparticles of the system across the partial gap in the spectrum near $k=0$. As a result they will have a threshold behavior depending on the size of the gap, but not necessarily on the mechanism by which it is created, or on the precise details of the quasiparticles of the system.

Single-electron correlations are also open to experimental measurement by tunnelling single electrons into the wire. If momentum is conserved, for instance in the case of tunneling between parallel wires, this allows a measurement of the spectral function $A(k,\omega)$.\cite{Jompol+2009,Schmidt+2012} On the other hand, both the d.c. conductance, and local tunneling of electrons from an STM tip give access to the tunneling density of states $\rho(\omega)$. These measurements are based on inserting single electrons into the system. Depending on the quantum numbers of the quasiparticles, tunneling of a single electron may require the creation of more than one quasiparticle, and so can lead to a different threshold behavior depending upon the gap-opening mechanism.

Whereas the quasiparticles near $k=0$ in the system with a $B$-field gap have a charge $e$, the same as the free electrons, we will show below that in the umklapp case the quasiparticles have charge $e/2$. So, to conserve charge in the tunnel process, it is necessary to create a \emph{pair} of quasiparticle excitations. This makes it possible to clearly see the difference between the competing gap-opening mechanisms of magnetic field and umklapp scattering. Moreover, the precise functional form of the response functions close to the threshold energy differ strongly between the two alternative gap-opening mechanisms.

Within this section, we calculate both the structure factor and the spectral function for a system with a partial gap generated either by a magnetic field or by umklapp scattering. We also compute the tunnelling density of states (TDOS) from the spectral function, to make direct contact with STM-based experiments. Since we are interested in the partially gapped phase, we will assume $\mu = 0$ and $B_y = 0$.

From here on, we choose to work in the $\pm$ basis for the bosonic modes. The Hamiltonian contains coupling terms between the modes ($\phi_-,\theta_-$) near $k = 0$ and the modes ($\phi_+,\theta_+$) near $k = \pm k_F$. One of these, the spin-flip term $\hat{V}_{\rm sf}$ is RG-irrelevant according to the previous section and can be neglected. The modes are also coupled by density-density interactions, which give rise to terms proportional to $(\partial_x \phi_-) (\partial_x \phi_+)$ and $(\partial_x \theta_-) (\partial_x \theta_+)$. To simplify our discussion, we start by neglecting these inter-mode couplings in Sec.~\ref{sec:FF}. Afterwards in Sec.~\ref{sec:FF_intermode}, we will briefly explain how to account for inter-mode couplings and show that they neither change the excitation threshold of the response functions, nor alter the qualitative difference between the magnetic field gap and the umklapp gap. However, they do generally change the exponents of the power-laws in response functions.

Hence, we first keep only the kinetic terms of the two uncoupled Luttinger liquids, but allow them to have different Luttinger parameters $K_\pm$ and sound velocities $v_\pm$ \, and the RG-relevant cosine term due to either a magnetic field, or umklapp scattering. The Hamiltonian we consider reads
\begin{align} \label{eq:solitonham}
\hat{H} &= \hat{H}_+ + \hat{H}_-, \\
\hat{H}_+&= \frac{v_+}{2 \pi} \int \, dx \, \left( K_+^{-1} (\partial_x \phi_+)^2 + K_+ (\partial_x \theta_+)^2 \right), \\
\hat{H}_-&= \frac{v_-}{2\pi} \int dx \left( K_-^{-1} (\partial_x \phi_-)^2 + K_- (\partial_x \theta_-)^2 \right) \nonumber \\
& + g_\gamma \int dx \cos (2 \gamma \phi_-). \label{eq:minusmodeham}
\end{align}
For $\gamma = 1$, the cosine term in $H_-$ describes a magnetic field perpendicular to the Rashba axis with $g_1=B_z/(\pi a)$, which is relevant for $K_-<2$. For $\gamma = 2$ it describes an umklapp scattering term with $g_2=g_\text{U}$, which becomes relevant as long as $K_-<1/2$.

The gapless outer modes have a standard Luttinger Hamiltonian, which decouples completely from the gapped inner modes. The contribution of the outer modes to response functions can then be calculated within the framework of Luttinger theory. The gapped central modes have a sine-Gordon Hamiltonian, and require a different approach based on the exact solubility of this sine-Gordon model.

To orient our discussion, we first set some notation and describe the response functions we will later calculate. The slow Fourier component of the total charge density for all the modes is $ \rho (x) = \sum_\sigma \, \rho_\sigma (x) = -\partial_x [\phi_+(x) + \phi_- (x) ]/\pi$. The density structure factor (\ref{eq:DSF}) for our simplified Hamiltonian (\ref{eq:minusmodeham}) becomes a sum of independent contributions from the $\phi_+$ and $\phi_-$ fields $S (k,\omega) = S_+(k,\omega) + S_-(k,\omega)$ where
\begin{equation} \label{eq:structurefactor2}
	S_\pm (k,\omega) = \frac{1}{\pi^2} \int \, dx dt e^{i\omega t-ikx} \langle \partial_x \phi_\pm(x,t) \partial_x \phi_\pm (0,0) \rangle.
\end{equation}
The spectral function for electrons with a given spin was defined in Eq.~(\ref{eq:SF}). Using a Lehman spectral representation, one can show that its particle part ($\omega>0$), corresponding to the insertion of an electronic state $\psi$ into the system is given by
\begin{equation}
A_\psi (k,\omega > 0) = \frac{1}{\pi} \text{Re} \int \, dx dt e^{i\omega t-ikx} \left\langle \psi (x,t) \psi^\dag (0,0) \right\rangle.
\end{equation}
With either cosine term present, spin is not conserved, and so the spectral function can receive a contribution from the gapped part of the Hamiltonian in Eq.~(\ref{eq:minusmodeham}).

\subsection{Soliton form factor solution}\label{sec:FF}

We use exact results from the integrability of the sine-Gordon model to calculate the behavior of the response functions, to find both the threshold behavior and the scaling of the response functions with $\omega$ just above their onset frequency.

The fundamental excitations of the sine-Gordon model (\ref{eq:minusmodeham}) are solitons, or kinks. The soliton creation operators, and their form factors are known in general for the sine-Gordon model [\onlinecite{Lukyanov+1997, Lukyanov+2001}]. Here, we use this information to investigate the spectral function and the structure factor of our model. An analogous procedure has been used to calculate the spectral function of a Hubbard model at commensurate filling.\cite{Tsvelik+2002}

We pass to the Lagrangian formalism, and rescale the fields to $\hat{\phi}_- = 2 \sqrt{2} \phi_-/\sqrt{K_-}$ and $\hat{\theta}_- = \sqrt{K_-} \theta_-/(2 \sqrt{2})$ to recover the standard form of the Lagrangian density
\begin{equation}
\mathcal{L} = \frac{1}{16 \pi} \left(\frac{1}{v_-}(\partial_t \hat{\phi}_-)^2 - v_-(\partial_x \hat{\phi}_-)^2 \right) + 2\zeta \cos(\beta \hat{\phi}_-),
\end{equation}
where $\zeta= g_\gamma/2$ and $\beta = \gamma \sqrt{K_-}/\sqrt{2}$. Note that the expressions we quote are valid for $0<\beta^2<1$, so apply only where the cosine is RG-relevant. In addition, the interaction between the solitons is known to be attractive for $0<\beta^2<1/2$, and repulsive for $1/2<\beta^2<1$. In the repulsive regime, we can build multi-soliton states from individual solitons, which are asymptotically free. In the regime in which the interactions are attractive, as well as multi-soliton states, we also find bound states of solitons known as breathers.

At $\beta^2=1/2$, the solitons are free. This point corresponds to $K_{LE}=1/\gamma^2$, which is called the Luther-Emery point of the theory, and is discussed in more detail below. This limit is tricky to approach within the soliton solution, a point we will elaborate on later, but it allows a solution based on refermionization.

Fundamental kink solutions are the lowest-energy excitations of the system, and interpolate between neighbouring minima of the cosine term as we go from $x=-\infty$ to $x=+\infty$. Sine-Gordon kinks have a relativistic dispersion relation $E(P) = (M_\gamma^2 + v_-^2 P^2)^{1/2}$, where up to dimensionless numerical factors $M_\gamma \propto g_\gamma^{1/[2(1-\beta^2)]}$ is the mass of the soliton.\cite{Zamolodchikov+1995} We can parametrize the energy and momentum of the kinks by their \emph{rapidity} $\theta = \text{arccosh} (1+v_-^2 P^2/M_\gamma^2)^{1/2}$ as
\begin{align}
E(\theta) = M_\gamma \cosh (\theta), \qquad
P(\theta) = \frac{M_\gamma}{v_-} \sinh (\theta). \label{eq:energymomsoliton}
\end{align}
In addition, the fundamental kinks also possess a $\mathbb{Z}_2$ charge $\epsilon=\pm1$, where $\epsilon=+1$ corresponds to a kink state, whereas $\epsilon=-1$ is its antiparticle, the \emph{anti-kink}.

An $N$-soliton state $|\Omega_N \rangle = |\theta_N,\epsilon_N,\theta_{N-1}, \epsilon_{N-1},\dots,\theta_1,\epsilon_1 \rangle$ can be built out of these fundamental kink states. The completeness relation for these $N$-soliton states is
\begin{align} \label{eq:setofstates}
\mathbb{I} &= \overset{\infty}{\underset{n=0}{\sum}} \underset{\epsilon_i}{\sum} \int_{-\infty}^{\infty} \frac{d\theta_1 \dots d\theta_n}{(2\pi)^n n!} |\Omega_n \rangle \langle \Omega_n |,
\end{align}
The vacuum state $|0\rangle$ is defined to contain no solitons.

\subsubsection{Density structure factor}

To calculate the density structure factor, we wish to use the form factors of the sine-Gordon model. We find from Ref.~[\onlinecite{Lukyanov+1997}] that the vertex operator $e^{is \hat{\phi}_-(x)}$ has a vacuum expectation value $\mathcal{G}_s = \langle 0 | e^{i s \hat{\phi}_- (x)} | 0 \rangle \approx 1 + \mathcal{O}(s^2)$. Taking derivatives gives $\langle 0| \hat{\phi}_- (x)|0 \rangle = \lim_{s \rightarrow 0} -i\partial_s \langle 0| e^{i s \hat{\phi}_- (x)}|0 \rangle \approx 0$, i.e., there is no contribution to the expectation value of the $\hat{\phi}_-$ operator in the zero-soliton sector. In fact, the leading contribution to the vertex function comes from a \emph{two soliton} state $|\Omega_2 \rangle = | \theta_2, \epsilon, \theta_1, -\epsilon \rangle$, which has the form factor \cite{Lukyanov+1997}
\begin{align}
\langle 0 | e^{i s \hat{\phi}_-(0)} | \Omega_2 \rangle = i s \epsilon \mathcal{G}_s F(\theta_1-\theta_2) , \label{eq:twosolitons}
\end{align}
where we have defined the function
\begin{equation}
F(\theta) = - \frac{G(\theta)}{G(-i\pi)} \frac{\pi}{\beta \cosh [\theta+i\pi/(2 \xi)] \cosh(\theta/2)},
\end{equation}
and the parameter $\xi = \beta^2/(1-\beta^2)$. $G(\theta)$ is given in integral representation as
\begin{widetext}
\begin{equation}
G(\theta) = i \sinh \left( \frac{\theta}{2}\right) \exp \left( \int_0^\infty \, \frac{dt}{t} \frac{\left\{ \sinh^2 [t \left(1-\frac{i \theta}{\pi} \right)] - \sinh^2 \left(\frac{t}{2}\right)\right\} \sinh[t(\xi-1)]}{\sinh(2t) \sinh(t\xi) \cosh(t)} \right).
\end{equation}
\end{widetext}

Taking a derivative of Eq.~(\ref{eq:twosolitons}) with respect to $s$, and using the translation operator, we find
\begin{align}
\langle 0 | \partial_x \hat{\phi}_-(x) | \Omega_2 \rangle = -\frac{2i \epsilon M_\gamma \pi}{\beta v_-} \sinh \left( \frac{\theta_1+\theta_2}{2} \right) \mathcal{G}_0 F(\theta_1 - \theta_2),
\end{align}
Since each soliton requires an energy of at least $M_\gamma$, such a state has an energy $E \geq 2M_\gamma$. Using the form factor, we can calculate the relevant correlation function by inserting a complete set of soliton states (\ref{eq:setofstates}). The leading contribution at low energies corresponds to states with at most two solitons, and reads
\begin{align}
\langle \partial_x \phi_-(x,t) \partial_x \phi_-(0) \rangle &= \frac{2 M_\gamma^2 \pi^2}{v_-^2 \gamma^2} \int \frac{d \theta_1 d \theta_2}{8\pi^2} e^{-i \overset{2}{\underset{j=1}{\sum}} [E(\theta_j)t + P(\theta_j)x]}\nonumber \\
&\times \sinh^2 \left( \frac{\theta_1+\theta_2}{2} \right) \mathcal{G}_0^2 |F(\theta_1 - \theta_2)|^2.
\end{align}
Fourier transforming this expression generates two $\delta$-functions, which correspond to energy and momentum conservation. The energy conservation $\delta$-function enforces that $S_-(k,\omega) = 0$ for $\omega < 2 M_\gamma$. To find the threshold behavior for
\begin{align}\label{eq:omega_th}
    \omega \gtrapprox \omega_{\rm th}(k) = \sqrt{4M_\gamma^2 + v_-^2 k^2} ,
\end{align}
we make the approximation that $F(\theta_1-\theta_2)^2 \sim f(K)^2 (\theta_1-\theta_2)^2$, which is true for small values of $\theta_1-\theta_2$, and $f(K)$ is a constant term which depends only on the value of the Luttinger parameter $K$. This approximation has a range of validity that depends on the Luttinger parameter through $K-K_{LE}$, i.e., on the distance from the Luther-Emery point. We find
\begin{align}\label{eq:Sminus}
S_-(k,\omega) \approx \frac{\pi^2 v_- k^2 f(K)^2}{4 \gamma^2 \sqrt{2} M_\gamma^3} \sqrt{\omega_{\rm th}(k) \, (\omega - \omega_{\rm th}(k))} \, \Theta [\omega-\omega_{\rm th}(k)], \notag \\
\end{align}
where $\Theta[\omega-\omega_{\rm th}(k)]$ is the Heaviside step function. Hence, near the threshold the density structure factor scales as $S_-(k,\omega) \propto \sqrt{\omega-\omega_{\rm th}(k)}$. Note that both the threshold, and the form of the density structure factor only depend on the value of $\gamma$ through the soliton mass, $M_\gamma$, otherwise the result is the same regardless of the mechanism responsible for the appearance of the gap.

At the Luther-Emery point $K = K_{LE}$, the approximation used in the paragraph after Eq.~(\ref{eq:omega_th}) breaks down. As we will show in Sec.~\ref{sec:LE}, one finds in this case the divergent threshold behavior $S_-(k,\omega) \propto 1/\sqrt{\omega-\omega_{\rm th}(k)}$, which coincides with the expected result for free fermions.

\begin{figure}[t]
	\centering
	\includegraphics[width=\columnwidth]{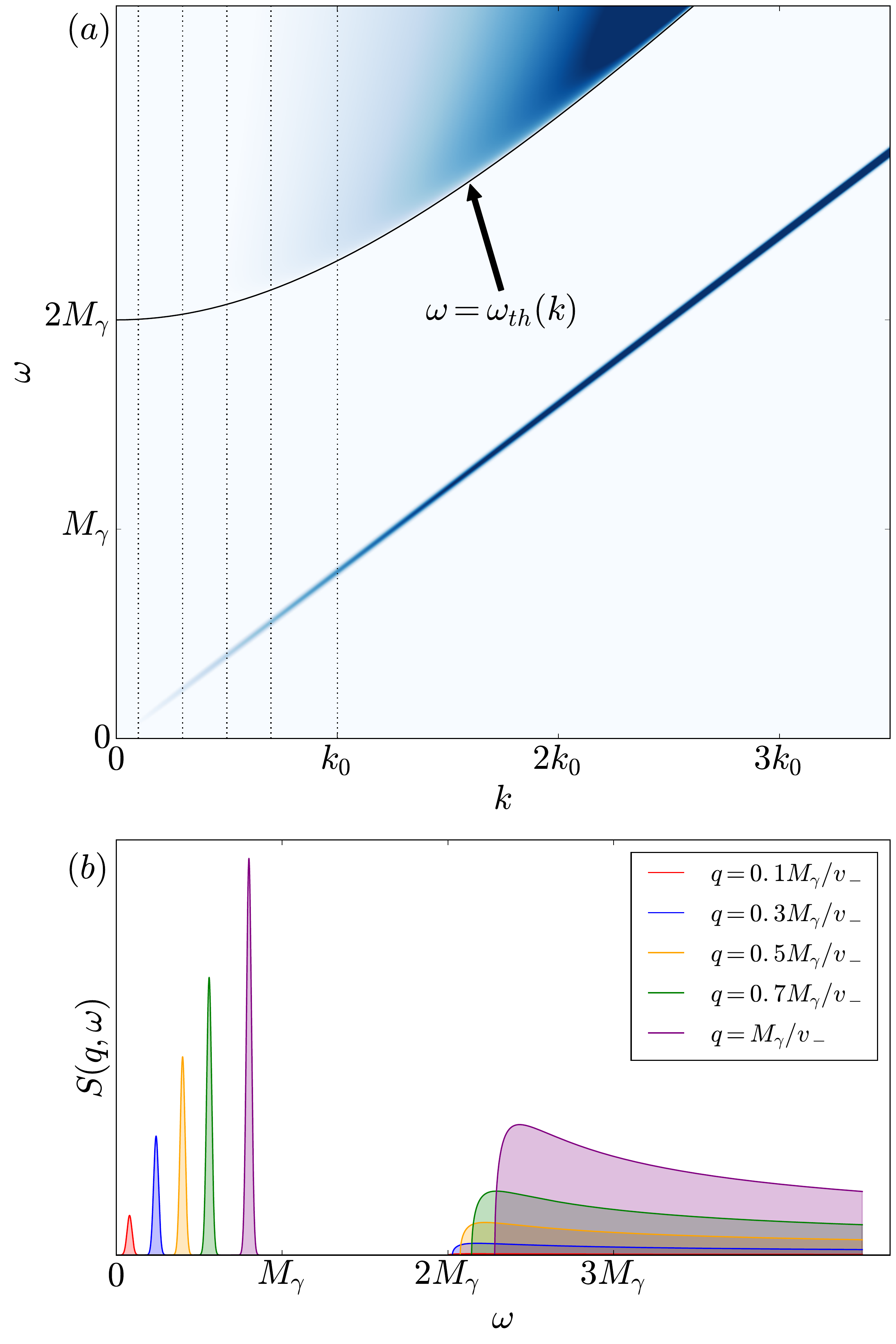}
	\caption{(Color online) (a) Density plot of the structure factor for the system with either a magnetic field or an umklapp-generated gap, encoded in the soliton mass $M_\gamma$. Regions with darker shading have a higher value of $S(k,\omega)$. The line through the origin is a smeared $\delta$-function peak arising from the Luttinger liquid modes, whereas the region of high $S(k,\omega)$ emanating from $\omega=2M_\gamma$ at $k=0$ corresponds to the contribution from the gapped modes. The momentum is measured in multiples of $k_0=M_\gamma/v_-$. The vertical dashed lines correspond to the cuts for fixed momentum of $S(k,\omega)$. (b) Plots of $S(q,\omega)$ for fixed momenta $q$. The size of the Luttinger liquid peaks grow linearly as a function of $k$, whereas the contribution from the gapped modes grow with $k^2$.}
	\label{fig:Dsfplot}
\end{figure}

We can calculate the contribution from the gapless outer modes, $\phi_+$ and $\theta_+$ within the framework of Luttinger theory. A straightforward calculation gives
\begin{equation}\label{eq:Splus}
S_+(k,\omega) = \frac{K_+ |k|}{4} \delta (\omega-v_+ k) ,
\end{equation}
which tells us that for a given value of $\omega$, we get a $\delta$-function spike at a finite value of $k$. This is quite different from the threshold behavior of $S_-(k,\omega)$, and even including thermal broadening effects will remain distinguishable from it. A plot of the threshold behavior of the density structure factor, including cuts at finite momentum is given in Fig.~\ref{fig:Dsfplot}.

Note that in the regime where $\beta^2<1/2$, the solitons attract one another, and we expect that breathing modes can also contribute to the density structure factor. The bound state resonances corresponding to these breathing modes will give $\delta$-function peaks in the structure factor at the dispersion relation of the breathing modes. Depending on the thermal or disorder broadening of these peaks, they may also be experimentally observable, and their number could be used to give bounds on the strength of interactions in the RSOC coupled wires.
\newline

\subsubsection{Spectral function}

In this section, we compute the spectral function $A_\psi(k,\omega)$, and the tunneling density of states for insertion of a particular electronic state into the wire. Specifically, we inject an electron with its spin in a general direction with respect to the Rashba axis $\psi = \cos(\theta/2) \psi_\uparrow + e^{i \phi} \sin(\theta/2) \psi_\downarrow$. As for the density structure factor, the spectral function factorizes into contributions from the gapped and the gapless modes. We first focus on the gapped modes, where we find
\begin{align}
A^-_\psi(k,\omega) &=    \frac{1}{2} \biggl[ A^-_{\uparrow \uparrow}(k,\omega) + A^-_{\downarrow \downarrow} (k,\omega) \biggr] \nonumber \\
& + \sin(\theta) \left[ e^{-i\phi}A^-_{\uparrow \downarrow}(k,\omega) +e^{i\phi} A^-_{\downarrow \uparrow}(k,\omega) \right],  \label{eq:specfntheta}
\end{align}
where the notation $A^-(k,\omega)$ denotes the contribution to the spectral function from the modes near to $k=0$, and
\begin{align}
A^-_{\uparrow \downarrow} (k,\omega) &= \frac{1}{\pi} \text{Re} \int dxdt e^{i \omega t - ikx} \langle \psi_{R \downarrow} (x,t) \psi_{L \uparrow}^\dag(0,0) \rangle, \label{eq:solitonspecfn} \\
A^-_{\uparrow \uparrow} (k,\omega) &= \frac{1}{\pi} \text{Re} \int dxdt e^{i \omega t - ikx} \langle \psi_{L \uparrow} (x,t) \psi_{L \uparrow}^\dag(0,0) \rangle.
\end{align}
These are the two quantities of interest (for both $\gamma=1$ and $\gamma=2$, we find $A^-_{\uparrow \uparrow}(k,\omega) = A^-_{\downarrow \downarrow}(k,\omega)$). We first focus on $A^-_{\uparrow \downarrow}(k,\omega) = A^-_{\downarrow \uparrow}(k,\omega)^\dag$. Using the space and time translation operators to rewrite $\psi_{R \downarrow}(x,t) = e^{i\hat{P} x} e^{i \hat{H} t} \psi_{R \downarrow} (0,0) e^{-i \hat{H} t} e^{-i \hat{P} x}$, and inserting a complete set of states shown in Eq.~(\ref{eq:setofstates}), we find
\begin{align}
\langle \psi_{R \downarrow} (x,t) \psi_{L \uparrow}^\dag(0,0) \rangle &= \overset{\infty}{\underset{n=0}{\sum}} \underset{\epsilon_i}{\sum} \int_{-\infty}^{\infty} \frac{d\theta_1 \dots d\theta_n}{(2\pi)^{n+1} a n!} e^{-i(Et+Px)} \nonumber \\
& \times \langle 0 | \psi_{R \downarrow}(0,0)|\Omega_n \rangle \langle \Omega_n | \psi_{L \uparrow}^\dag(0,0) | 0 \rangle.
\end{align}

Inserting the bosonized expressions for $\psi_{R \downarrow}(0,0)$ and $\psi_{L \uparrow}(0,0)$, we find we want to calculate $\langle 0 |e^{i(\pm \beta \hat{\phi}_-/2\gamma + \gamma \hat{\theta}_-/4\beta )} | \Omega_n \rangle$. We take only the lowest energy configuration which contributes to this expectation value, which has a form factor
\begin{align}
&\langle 0 | e^{2i \gamma \hat{\theta}_-/\beta} e^{\pm i \beta \hat{\phi}_-/2 \gamma} | \theta_n,\epsilon_n =-1,\dots,\theta_1, \epsilon_1=-1 \rangle  \nonumber \\
&= \sqrt{Z_\gamma(\beta/(2\gamma))} e^{\pm i\pi/4} \overset{\gamma}{\underset{j=1}{\prod}} e^{\pm \theta_j/2\gamma} \overset{n}{\underset{j<j^\prime}{\prod}} G(\theta_{j^\prime}-\theta_j).
\end{align}
where $Z_\gamma (s)$ is a complicated normalization factor which does not change the threshold behavior of our response functions, and is given explicitly in Ref.~[\onlinecite{Lukyanov+2001}].

Substituting this into Eq.~(\ref{eq:solitonspecfn}), and doing the integrals over $t$ and $x$, which give energy- and momentum-conservation $\delta$-functions, we recover
\begin{widetext}
	\begin{align} \label{eq:solitonspecfn2}
	A^-_{\uparrow \downarrow}(k,\omega) &= -\frac{i Z_\gamma(\beta/2 \gamma)}{(2\pi)^{\gamma-1} \pi a \gamma!} \int_{-\infty}^{\infty} d\theta_1 \dots d\theta_\gamma \overset{\gamma}{\underset{j<j^\prime}{\prod}} |G(\theta_{j^\prime} - \theta_j)|^2 \delta \left(k + \frac{M_\gamma}{v_-} \overset{\gamma}{\underset{i=1}{\sum}} \sinh (\theta_i) \right) \delta \left( \omega - M_\gamma \overset{\gamma}{\underset{i=1}{\sum}} \cosh (\theta_i) \right).
	\end{align}
\end{widetext}
Note that the summations and products in these expressions run over values which depend on $\gamma$. However, at this point, we can already see the onset value of the spectral function. The function $\cosh(\theta) \geq 1$, so in order to satisfy the energy conservation $\delta$-function, we must have $\omega > \gamma M_\gamma$. Note that the threshold, and the form of the spectral function depend \emph{explictly} on the value of $\gamma$, as well as through the soliton mass, $M_\gamma$.

We calculate the form of the spectral function exactly for the different gap mechanisms. For $\gamma=1$, the expression Eq.~(\ref{eq:solitonspecfn2}) simplifies to
\begin{align}
A^-_{\uparrow \downarrow} (k,\omega) = \frac{-i Z_1(\beta/2)}{\pi a} \int_{-\infty}^{\infty} d\theta &\delta [k + \frac{M_1}{v_-} \sinh (\theta)] \notag \\
& \times \delta [\omega - M_1 \cosh (\theta)].
\end{align}
We can do this integral exactly to find
\begin{equation}\label{eq:Aminus_updown}
A^-_{\uparrow \downarrow} (k,\omega) =  \frac{C_1}{\omega_{{\rm th},1}(k)} \delta (\omega - \omega_{\rm th,1}(k)),
\end{equation}
where the constant term $C_1=-i v_- Z_1(\beta/2)/(\pi a)$, and  $\omega_{{\rm th},1}(k) = [M_1 +v_-^2k^2]^{1/2}$. We therefore get a $\delta$-function singularity in the spectral function at $\omega = \omega_{\rm th,1}(k)$. A similar calculation for $A^-_{\uparrow \uparrow} (k,\omega)$ gives the result
\begin{equation}
A^-_{\uparrow \uparrow} (k,\omega) = \tilde{C}_1 \left[ 1-\frac{v_- k}{\omega_{{\rm th},1}(k)} \right] \delta (\omega - \omega_{{\rm th},1}(k)),
\end{equation}
where $\tilde{C}_1= v_- Z_1(\beta/2)/(\pi a)$. With these two functions, we proceed to calculate the contribution to the TDOS from the gapped modes $\rho^-_\psi(\omega)$ for a general spin-polarized state $\psi$ from Eq.~(\ref{eq:specfntheta}), simply by integrating over $k$. We find
\begin{align}
\rho^-_\psi(\omega) = \frac{2Z_1(\beta/2)}{\pi a M_1 \sqrt{\omega^2 - M_1^2}} \left[ \omega + M_1 \sin (\theta) \sin (\phi) \right] \Theta(\omega-M_1).
\end{align}
Just above the threshold, $\omega= M_1 + \delta \omega$, this expression consists of two contributions. The first is an angle-independent background which behaves like $1/\sqrt{\delta \omega}$, and an angularly varying term which is proportional to $\sin(\theta)\sin(\phi)$, and also scales like $1/\sqrt{\delta \omega}$. This $1/\sqrt{\delta \omega}$ behavior can be understood as the usual van Hove singularity for a one-dimensional system.

Note that Ref.~[\onlinecite{Schuricht+2012}] considers a similar Hamiltonian to Eq.~(\ref{eq:solitonham}) with $\gamma=1$. However, their definition of the fermionic fields in terms of the bosonic fields is different to ours, so that when they make the approximation of discarding marginal inter-mode interaction terms, they are discarding terms of a different physical origin to us. As a result, whilst our results agree with those of Ref.~[\onlinecite{Schuricht+2012}] for the threshold energy for the spectral function, since we are discarding different marginal terms, we find different power law behavior to them above the onset.

\begin{figure}[t]
	\centering
	\includegraphics[width=\columnwidth]{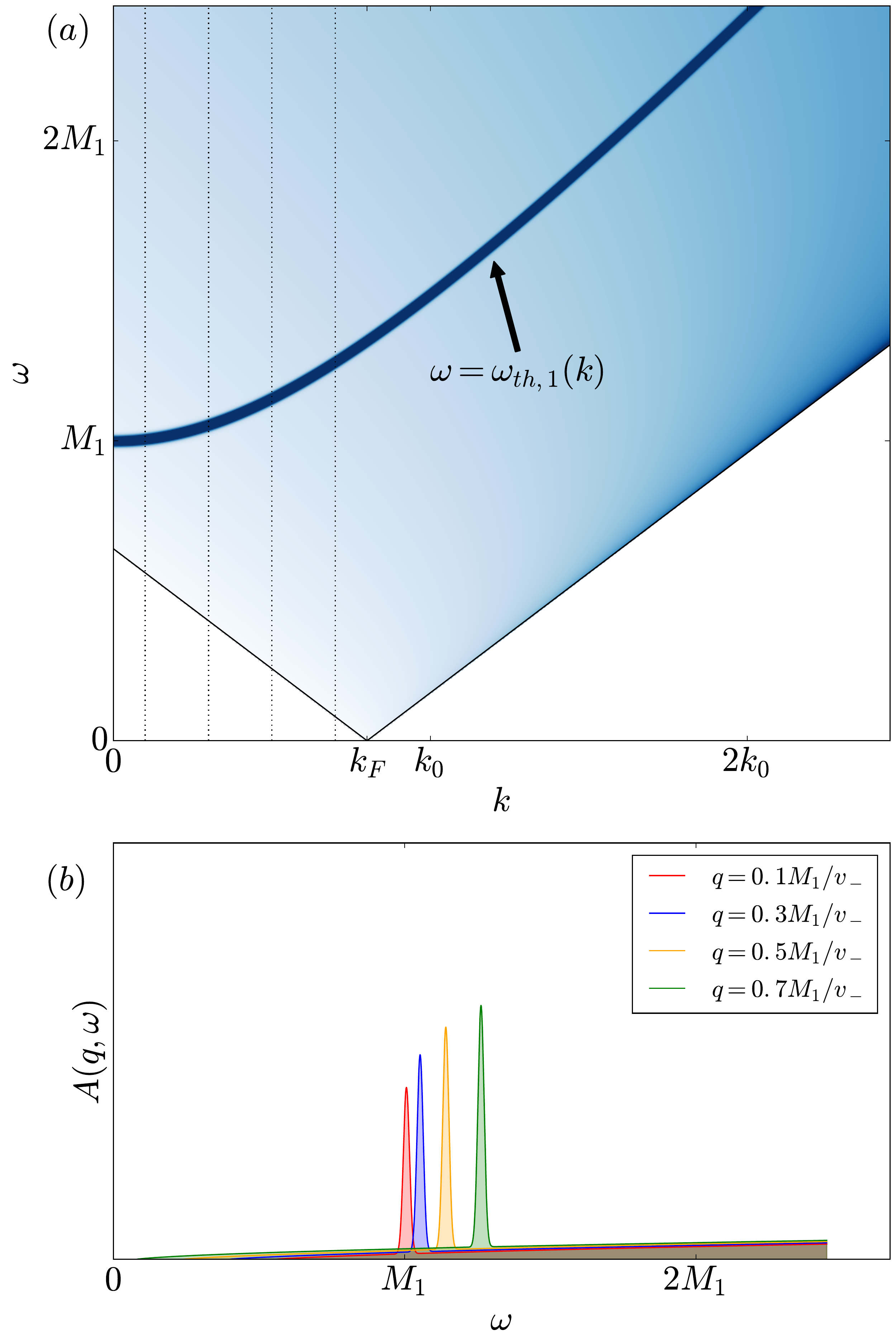}
	\caption{(Color online) (a) Density plot of the spectral function with a magnetic field induced gap, whose size is encoded in the soliton mass $M_1$. Regions with darker shading have a higher value of $A(k,\omega)$. The fan of spectral weight spreading from $k_F = 2m \alpha_R$ comes from the Luttinger liquid modes, whereas the smeared $\delta$-function peak emanating from $\omega=M_1$ at $k=0$ corresponds to the contribution from the gapped modes. The momentum is measured in multiples of $k_0=M_1/v_-$. The vertical dashed lines correspond to the cuts for fixed momentum of $A(k,\omega)$. (b) Plots of $A(q,\omega)$ for fixed momenta $q$. The Luttinger liquid contribution is broad and featureless for momenta $q<k_F$, whereas the contribution from the gapped modes gives a sharp peak.}
	\label{fig:sfplotb}
\end{figure}

For $\gamma=2$, the spectral function is more complicated. The $A^-_{\uparrow \downarrow}(k,\omega)$ contribution takes the form
\begin{widetext}
	\begin{align}
	A^-_{\uparrow \downarrow} (k,\omega) &= -\frac{i Z_2(\beta/4)}{(2\pi)^2 a} \int_{-\infty}^{\infty} d\theta_1 d\theta_2 |G(\theta_1 - \theta_2)|^2 \delta \left(k + \frac{M_2}{v_-} ( \sinh (\theta_1) + \sinh (\theta_2)) \right) \delta \left( \omega - M_2 (\cosh (\theta_1) + \cosh (\theta_2)) \right),
	\end{align}
\end{widetext}
We make the approximation that $|G(\theta_1-\theta_2)|^2 \sim g(K)^2 (\theta_1-\theta_2)^2$, where $g(K)$ is again a constant which depends only on the Luttinger $K$, and which is valid for small values of $\theta_1-\theta_2$. The result is that
\begin{align}\label{eq:Aminus_updown_umklapp}
A^-_{\uparrow \downarrow} (k,\omega) = -\frac{i v_- Z_2(\beta/4) g(K)^2}{(2\pi)^2 a} &\frac{\sqrt{\omega_{{\rm th},2}(k)[\omega-\omega_{{\rm th},2}(k)]}}{\sqrt{2}M_2^3} \notag \\
& \times \Theta[\omega-\omega_{{\rm th},2}(k)],
\end{align}
where we have defined $\omega_{{\rm th},2}(k) = [4M_2^2 + v_-^2 k^2]^{1/2}$. We see that for small momenta, the spectral function switches on with a square-root cusp at energy $\omega \sim 2M_2$. A similar calculation for $A^-_{\uparrow \uparrow}(k,\omega)$ shows that in the case of an umklapp gap,
\begin{align}
A^-_{\uparrow \uparrow}(k,\omega) &= \frac{v_- Z_2(\beta/4)}{2\sqrt{2} (2\pi)^2  a M_2^4} \left[1-\frac{v_- k}{\sqrt{4M_2^2 + v_-^2 k^2}} \right] \notag \\
& \hspace{5mm} \times [\omega_{{\rm th},2}(k)]^{\frac{3}{2}} \sqrt{\omega-\omega_{{\rm th},2}(k)} \, \Theta[\omega-\omega_{{\rm th},2}(k)].
\end{align}
In this case, the TDOS for the injection of a spin polarized state gets a contribution from $A_{\uparrow \downarrow}^-(k,\omega)$ which takes the form
\begin{equation}
\rho^-_{\uparrow \downarrow} (\omega) =  \frac{Z_2(\beta/4) g(K)^2}{(2\pi)^2 a \sqrt{2}M_2^3} \sin \theta \sin \theta \int_{2M_2}^{\omega}  du \, \frac{u^{\frac{3}{2}} \sqrt{\omega - u}}{\sqrt{u^2-4M_2^2}}.
\end{equation}
For small $\omega$, we approximate the integral to find
\begin{equation}
\rho^-_{\uparrow \downarrow} =  \frac{Z_2(\beta/4) g(K)^2}{8 \pi a M_2^2 } (\omega-2M_2) \,\Theta(\omega-2M_2).
\end{equation}
A similar calculation for the leading contribution from $A_{\uparrow \uparrow}(k,\omega)$ gives that
\begin{equation}
\rho^-_{\uparrow \uparrow} = \frac{Z_2(\beta/4)}{2\sqrt{2}(2\pi)^2 a M_2^4} \int_{2M_2}^{\omega}  du \, \frac{u^{\frac{5}{2}} \sqrt{\omega - u}}{\sqrt{u^2-4M_2^2}}.
\end{equation}

This result tells us that in the case of an umklapp-generated gap, we have a linear behavior of the tunnelling density of states with energy for values just above the threshold. For the injection of a general spin-polarized state, we find that
\begin{equation}
\rho_\psi^-(\omega) \approx \Gamma \left\{1+\sin \theta \sin \phi \right\} (\omega - 2M_2) \Theta (\omega - 2M_2)
\end{equation}
where $\Gamma = Z_2(\beta/4)g(K)^2/(8\pi a M_2^2)$. Above the threshold value, $\omega = 2M + \delta \omega$, we find that the TDOS is linear in $\delta \omega$, in contrast to the magnetic field-induced gap it behaves as $1/\sqrt{\delta \omega}$ for energies just above the threshold.

\begin{figure}[t]
	\centering
	\includegraphics[width=\columnwidth]{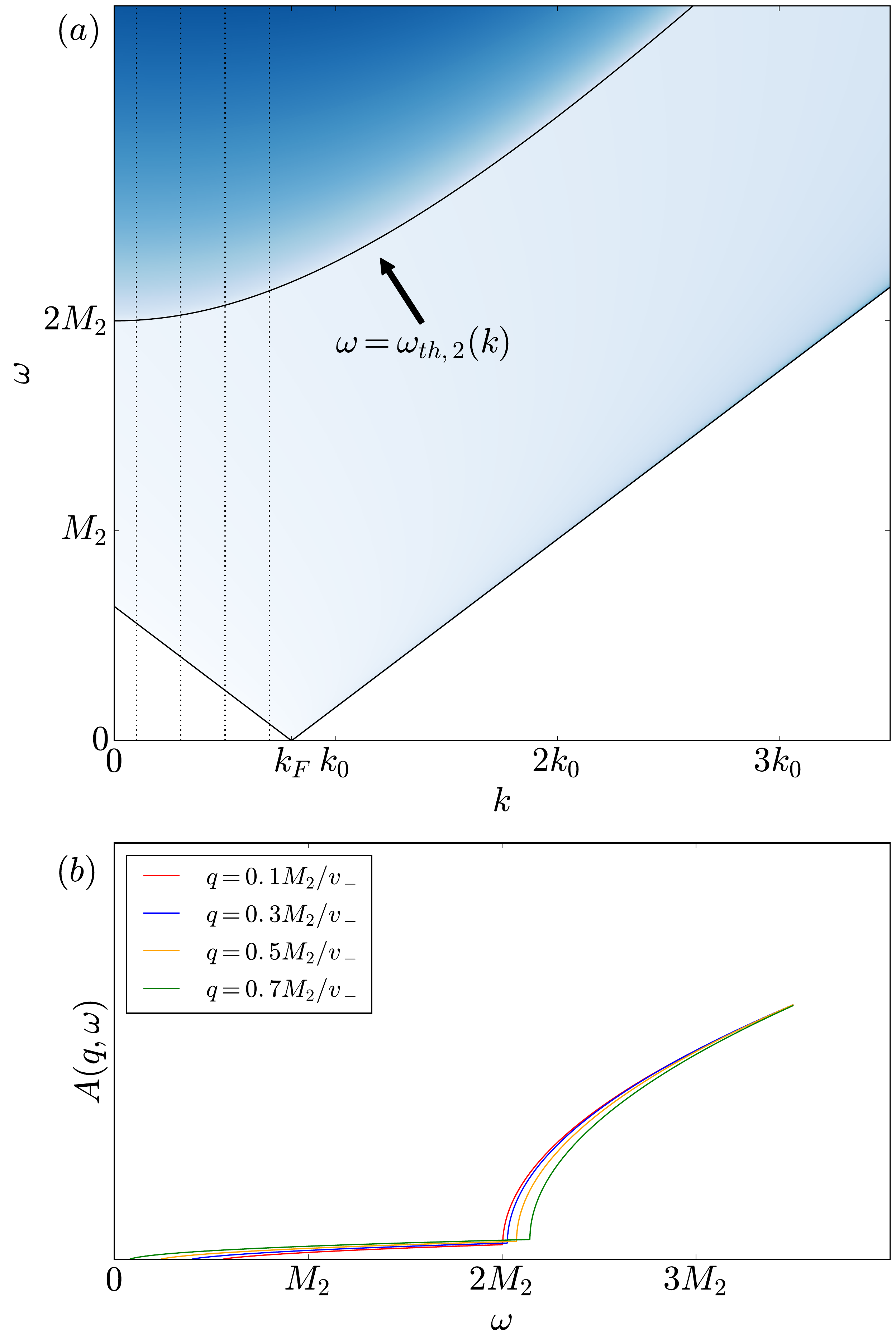}
	\caption{(Color online) (a) Density plot of the spectral function with an umklapp gap, whose size is encoded in the soliton mass $M_2$. Regions with darker shading have a higher value of $A(k,\omega)$. The fan of spectral weight spreading from $k_F = 2m \alpha_R$ comes from the Luttinger liquid modes, whereas the square-root function which onsets near to $\omega=2M_2$ is the contribution from the gapped modes. The momentum is measured in multiples of $k_0=M_2/v_-$. The vertical dashed lines correspond to the cuts for fixed momentum of $A(k,\omega)$. (b) Plots of $A(q,\omega)$ for fixed momenta $q$. The Luttinger liquid contribution is broad and featureless for momenta $q<k_F$, whereas the contribution from the gapped modes gives a square-root onset.}
	\label{fig:sfplotum}
\end{figure}

The contribution from the ungapped, outer modes can again be calculated within the framework of Luttinger theory, with the result that
\begin{equation}
A^+_\psi(k,\omega) = C_+ \prod_{\alpha,\beta=\pm} \left[\frac{\omega+ \alpha v_+k}{2v_+}\right]^{\delta_\alpha-1}  \Theta[w+\alpha v_+(k+\beta k_F)]
\end{equation}
where $k_F = 2m\alpha_R$, the exponents $\delta_\alpha = (K_+ +1/K_+ + 2\alpha)/4$ depend on the Luttinger parameter $K$, and $C_+$ is a non-universal number which depends on the cutoff $a$. For $K_+\ll 1$, $\delta_\alpha - 1 >0$, and this term gives a pair of fan-like contributions emanating from $k=\pm k_F$ (see Fig.~\ref{fig:sfplotb} and Fig.~\ref{fig:sfplotum} ).

\subsection{Response functions with inter-mode coupling}\label{sec:FF_intermode}

In Sec.~\ref{sec:FF}, we calculated the response function based on the Hamiltonian (\ref{eq:solitonham})-(\ref{eq:minusmodeham}), in which we neglected coupling terms between the modes near $k=0$ and near $k=\pm k_F$. As we have shown in Eq.~(\ref{eq:Vrho}), the interaction processes producing these terms require a momentum transfer of $k_F$ and are thus small for an interaction potential with nonzero range, for which $\tilde{V}(k_F) < \tilde{V}(0)$. Despite their smallness, however, we show in this sections that these coupling terms can change the response functions quantitatively.

The inter-mode coupling terms arising from Eq.~(\ref{eq:Vrho}) can be written as
\begin{align}
    H_{+-} = \int dx \left[ g_\theta (\partial_x  \theta_+) (\partial_x  \theta_-) + g_\phi (\partial_x  \phi_+) (\partial_x  \phi_-) \right].
\end{align}
When the cosine term in Eq.~(\ref{eq:minusmodeham}) becomes relevant, it tends to pin the $\phi_-$ field, so the coupling term proportional to $g_\phi$ is strongly suppressed. In contrast, however, a pinned $\phi_-$ entails strong fluctuations in its conjugate field $\theta_-$, so the $g_{\theta}$ term should in principle be taken into account.

Let us briefly discuss how to do this, and to simplify the algebra let us assume equal Fermi velocities $v_+ = v_- = v$ and Luttinger parameters $K_+ = K_- = K$. Shifting two of the four bosonic fields by $\theta_- \to \theta_- + \alpha \theta_+$ and $\phi_+ \to \phi_+ - \alpha \phi_-$, one can eliminate the $g_\theta$ coupling term by choosing $\alpha = \pi g_\theta/(K v)$. The transformation leaves us with a Hamiltonian of the form (\ref{eq:solitonham})-(\ref{eq:minusmodeham}) with renormalized parameters, supplemented with a $g_\phi$ coupling term, but the latter can be neglected if the cosine is relevant.

The transformation used to eliminate the $g_\theta$ term causes changes in the response functions. In the density structure factor $S(k,\omega)$, it merely changes the prefactors in Eqs.~(\ref{eq:Sminus}) and (\ref{eq:Splus}). For the spectral function $A(k,\omega)$, in contrast, the changes are more significant because the shifts in bosonic fields also change the bosonization formula for the fermionic operators. Most importantly, the $k\approx 0$ fields change into
\begin{align}\label{eq:psi_shift}
    \psi_{R\downarrow} \to e^{-i (\phi_- - \theta_- + \alpha \theta_+)},\quad
    \psi_{L\uparrow} \to e^{-i (-\phi_- - \theta_- + \alpha \theta_+)}.
\end{align}
The shift (\ref{eq:psi_shift}) strongly affects the both the spin-flip contributions $A^-_{\uparrow\downarrow}$, $A^-_{\downarrow\uparrow}$ and the equal-spin contributions $A^-_{\uparrow\uparrow}$ and $A^-_{\downarrow\downarrow}$ to the spectral function. These experience a change in scaling resulting from the inter-mode coupling. This coupling allows a part of the energy of the injected particle near $k=0$ to be used to create particle pairs with opposite spins at the Fermi points $\pm k_F$ because $e^{i \alpha \theta_+} \propto [\psi_{L\downarrow}^\dag \psi^\dag_{R\uparrow}]^{\alpha/2}$. A lengthy but straightforward calculation shows that this changes the function $A^-_{\uparrow\downarrow}$ from a Dirac-$\delta$ function to a power-law singularity. Instead of Eq.~(\ref{eq:Aminus_updown}), one now finds for a $B$-field generated gap,
\begin{align}
    A^-_{\uparrow \downarrow} (k,\omega) \propto [\omega - \omega_{\rm th,1}(k)]^{\alpha^2/2-1},
\end{align}
where for the case of an umklapp gap, Eq.~(\ref{eq:Aminus_updown_umklapp}) is replaced by
\begin{align}
    A^-_{\uparrow \downarrow} (k,\omega) \propto [\omega - \omega_{\rm th,2}(k)]^{(\alpha^2+1)/2}.
\end{align}
The limit $\alpha \to 0$ leads back to a representation of the $\delta$-function for the magnetic field gap, and a square root cusp for the umklapp gap, respectively.

\subsection{Luther-Emery solution}\label{sec:LE}

In order to get a physical picture of our soliton solutions, and also to investigate the free soliton limit, which is not amenable to the approximations we made in the soliton computation, we carry out a parallel calculation of the density structure factor by a different approach.

The central idea of Luther and Emery [\onlinecite{Emery+1974}] was that a sine-Gordon Hamiltonian can correspond to a theory of \emph{free} quasiparticles for a particular, special value $K=K_{LE}$. At this \emph{Luther-Emery} point, we may re-fermionize Eq.~(\ref{eq:minusmodeham}) in terms of new quasiparticle operators, which are different from the original fermions. The value of $K_{LE}$ depends on the gap-opening mechanism $\gamma$ through $K_{LE} = 1/\gamma^2$. For both $\gamma=1$ and $\gamma=2$, this corresponds to a regime where the cosine is relevant.

We introduce new fermion operators according to
\begin{equation} \label{eq:newfermions}
\bar{\psi}_{\alpha} (x) = \frac{1}{\sqrt{2 \pi a}} e^{-i(\alpha \bar{\phi} (x) -\bar{\theta}(x)) },
\end{equation}
where $\bar{\theta} = \sqrt{K_{LE}} \theta_-$ and $\bar{\phi} = \phi_-/\sqrt{K_{LE}}$, and we have dropped the minus subscripts on all fields. As usual, $\alpha=R,L$. We can express these field operators in terms of the quasiparticles for the different values of $\gamma$.

In the case $\gamma=1$, where $K_{LE}=1$, we have $\bar{\phi}=\phi_-$ and $\bar{\theta}=\theta_-$, so refermionization is trivial; the ``new'' fermions $\bar{\psi}$ are really the original degrees of freedom $\psi$.

For $\gamma=2$, $K_{LE}=1/4$ so that $\bar{\theta} = \theta_-/2$ and $\bar{\phi} = 2\phi_-$. For the fermions, we find
\begin{equation}
\bar{\psi}_{\alpha} (x) = \frac{1}{\sqrt{2 \pi a}} e^{-i(2\alpha {\phi}_- (x) -\theta_-(x)/2) }.
\end{equation}
Writing the original fermions in terms of these new quasiparticles, we write
\begin{align}
\psi_{R \downarrow} (x) &= \frac{1}{\sqrt{2 \pi a}} e^{-i (\bar{\phi}/2 -2\bar{\theta})} \nonumber \\
&=  \frac{1}{\sqrt{2 \pi a}} e^{-2i (\bar{\phi} - \bar{\theta})} e^{3i \bar{\phi}/2} \nonumber \\
&\propto  (2 \pi a)^{3/2} \bar{\psi}_{R} \partial_x \bar{\psi}_{R} e^{(3\pi i/2) \int_{-\infty}^{x} (\bar{\rho}_{L} + \bar{\rho}_{R} )}.
\end{align}
Since the densities under the integral come from the gapped modes, at low energies we can assume that the string operator in the final line is approximately unity. We find
\begin{align} \label{eq:gamma2op}
\psi_{\alpha} (x) & \approx (2 \pi a)^{3/2} \bar{\psi}_{\alpha}\partial_x \bar{\psi}_{\alpha},
\end{align}
so a single electron excitation maps to \emph{two} quasiparticles. In both cases, going to momentum space, Eq.~(\ref{eq:minusmodeham}) becomes
\begin{align}
H_- &= \underset{k}{\sum} \left( \begin{array}{c c } \bar{\psi}_{R,k}^\dag & \bar{\psi}_{L,k}^\dag \end{array} \right) \left( \begin{array}{c c} v_- k & M_\gamma \\ M_\gamma & -v_- k \end{array} \right) \left( \begin{array}{c} \bar{\psi}_{R,k} \\ \bar{\psi}_{L,k} \end{array} \right),
\end{align}
where we have defined $M_1 = B_z$ and $M_2 = \pi a g_{\text{U}}$. Rotating the basis of operators according to
\begin{equation} \label{eq:rotation}
\left( \begin{array}{c} \bar{\psi}_{R,k} \\ \bar{\psi}_{L,k} \end{array} \right) = B_k^\gamma \left( \begin{array}{c} c_{R,k} \\ c_{L,k} \end{array} \right) \quad, \quad B_k^\gamma = \left( \begin{array}{c c} \cos\zeta(k) & -\sin \zeta(k) \\ \sin \zeta (k) & \cos \zeta(k) \end{array} \right),
\end{equation}
where $\zeta^\gamma(k) = \pi/4 - \arctan (v_- k/M_\gamma)/2$, we can diagonalize this Hamiltonian exactly to find
\begin{align}
H_- &= \underset{k,\alpha}{\sum} \xi_{\alpha,k}^\gamma c^\dag_{\alpha,k} c_{\alpha,k},
\end{align}
where $\xi_{\alpha,k}^\gamma = \alpha [(v_- k)^2 + M_\gamma^2]^{1/2}$. Straight away we can see that the single quasiparticle spectrum has a spectral gap of size $2 M_\gamma$ at $k=0$. What does this mean for the real electrons? To find out, we calculate the density structure factor in terms of these re-fermionized quasiparticle modes.

We focus on the contribution $S_-(k,\omega)$ to Eq.~(\ref{eq:structurefactor2}) from the central, gapped modes, which is given by
\begin{align}
S_- (k,\omega) &= \frac{1}{\pi^2 \gamma^2} \int \, dx dt e^{i\omega t-ikx} \langle \partial_x \bar{\phi}_-(x,t) \partial_x \bar{\phi}_- (0,0) \rangle \nonumber \\
& \hspace{-6mm} = \frac{1}{\gamma^2 L} \underset{\alpha,\alpha^\prime, q}{\sum} \int \, dt e^{i\omega t} \langle \bar{\psi}_{\alpha,q}(t) \bar{\psi}_{\alpha^\prime,q}^\dag (0)\rangle \langle \bar{\psi}_{\alpha,q-k}^\dag(t) \bar{\psi}_{\alpha^\prime,q-k}(0)\rangle.
\end{align}

Rotating to the basis of operators which diagonalizes the Hamiltonian, given in Eq.~(\ref{eq:rotation}), we find that
\begin{equation}
S_-(k,\omega) = \frac{1}{\gamma^2 L} \underset{q}{\sum} \delta [\omega - \xi_{-,q}^\gamma - \xi_{-,q-k}^\gamma] \sin^2 (\zeta(q-k)-\zeta(q)),
\end{equation}
where the delta function imposes energy conservation, and the sine function accounts for the overlap of the initial and final states. Re-writing the delta function as a constraint on $q$, we find that for the equation $\omega - \xi_{-,q}^\gamma - \xi_{-,q-k}^\gamma$ to have real solutions $q_i$ for $q$, we must have $\omega >  \tilde{\omega}_{\rm th}$, where
\begin{equation}
\tilde{\omega}_{\rm th} = \sqrt{v_-^2 k^2 + 4M_\gamma^2}.
\end{equation}
In other words, $S_-(k\approx 0,\omega) = 0$ if $\omega < 2M_\gamma$. The exact form of the gap opening can be calculated, and we find
\begin{equation}
S_-(k\approx 0,\omega) = \frac{v_-^2 k^2}{2 M_\gamma^2}\frac{\Theta(\omega - \tilde{\omega}_{\rm th})}{4 \gamma^2 \pi v_-} \sqrt{\frac{M_\gamma}{\omega-\tilde{\omega}_{\rm th}(k)}}.
\end{equation}
This result tells us that we must excite a quasiparticle across the band gap to get a non-zero result for the density structure factor, regardless of the mechanism by which the spectral gap opens. This result agrees with that found from the soliton form factor calculation both for the threshold at which we find $S \neq 0$. However, the functional form of the density structure factor above the threshold has a $1/\sqrt{\omega-\tilde{\omega}_{\rm th}(k)}$ singularity, and not a $\sqrt{\omega-\tilde{\omega}_{\rm th}(k)}$ cusp as found from the soliton solution.

We can in fact connect this result with those we found from the soliton computation. Moving away from the Luther-Emery point $K=K_{LE}$ in our theory, we would then have to deal with interactions between the quasiparticles. In Ref.~[\onlinecite{Gangadharaiah+2013}], the authors compute the structure factor for a weakly-interacting helical edge state with a magnetic field by re-summation of an infinite series of ladder diagrams. The authors focus on inter-band transitions, which take quasiparticles from the filled lower band into the empty upper band, as these are the transitions most susceptible to interactions. Including a contact interaction $\tilde{V}$ between the quasiparticles, the structure factor is found to be
\begin{align}
S_-(k,\omega) &\propto \frac{k^2}{M_\gamma^{3/2}} \frac{\sqrt{\omega-\tilde{\omega}_{\rm th}}}{\omega - \tilde{\omega}_{\rm th} + \pi^2 \tilde{V}M_\gamma^2} \Theta (\omega - \tilde{\omega}_{\rm th}) \nonumber \\
&+ \frac{k^2 \tilde{V}}{M_\gamma} \delta (\omega - \tilde{\omega}_{\rm th} + \pi^2 M_\gamma \tilde{V} ) , \label{eq:suhasetal}
\end{align}
$\tilde{\omega}_{\rm th}$ is the lower threshold for a non-zero contribution which in turn depends on the value of $M_\gamma$, which is the size of the induced gap. The second term in Eq.~(\ref{eq:suhasetal}) comes from a bound state of quasiparticles, a breathing mode in our language, and will concern us less. The first term crosses over from having a $1/\sqrt{\omega-\tilde{\omega}_{\rm th}(k)}$ divergence when the fermions are free (i.e. $\tilde{V} = 0$), which matches our Luther-Emery calculation, to having a $\sqrt{\omega-\tilde{\omega}_{\rm th}(k)}$ cusp when $\tilde{V} \neq 0$ in agreement with our soliton calculation.

\section{Conclusions} \label{sec:conclusions}

We have discussed how a helical gap can arise in multi-subband quantum wires with Rashba spin-orbit coupling, due to an applied magnetic field, or umklapp scattering. This spin-non-conserving umklapp process is generated through a combination of interactions, and spin-flipping transitions to higher sub bands of the confining potential. For sufficiently strong interactions, and chemical potential near to the band crossing, this can open a similar gap to that arising from a perpendicular magnetic field, but in the absence of time reversal symmetry-breaking perturbations.

The opening of a helical gap would be visible to e.g.~conductance measurements. We point out that physical measurements may also be able to discriminate between a magnetic field generated partial gap, and one arising from umklapp scattering. In particular, we find that the ratio of the threshold energies for onset of the spectral function and the density structure factor is one for the magnetic field gap, but is two for the umklapp gap, due to the fractionalized charge of the quasiparticles in the latter case. The onset of these functions occurs in a region of $(k,\omega)$ space where we expect a generically featureless contribution from the gapless outer branches.

The tunneling density of states is also very different between the two cases. In the case of a magnetic field generated helical gap, at frequencies just above the threshold $\omega = M_1 + \delta \omega$, the TDOS switches on with a van Hove-like $1/\sqrt{\delta \omega}$ dependence. In contrast, when umklapp scattering causes the helical gap, we find that the TDOS depends linearly on $\delta \omega = \omega -2M_2$. 

We would like to comment briefly on the relation between the two-dimensional rectangular geometry we have chosen in Eq.~(\ref{RashbaHamiltonian}) and experimental InAs and InSb wires. Firstly, even though these wires usually have a hexagonal cross section, they can be well approximated as cylindrical. In InSb, electrons are accommodated close to the core of the wire, and so are only weakly sensitive to the hexagonal symmetry of the wire's surface.\cite{Kammhuber+2016} In InAs, on the other hand, Fermi-level pinning causes electronic accumulation nearer to the surface, but even in this case, experimental measurements agree well with models based on cylindrical wires.\cite{hernandez+2010}

Secondly, for a Rashba wire with cylindrical cross section, the subbands can be classified as eigenstates of the total angular momentum operator. In that case, the ground state has a twofold spin degeneracy analogous to our model in Eq.~(\ref{RashbaHamiltonian}). The first excited state, in contrast, has a twofold spin and a twofold orbital degeneracy, the latter corresponding to clockwise and counterclockwise electron motion, in contrast to our model, whose first excited state is only spin-degenerate.

However, even for cylindrical wires, it is possible to show that transitions between subbands are associated with a spin flip as in Eq.~(\ref{H1}). Hence, within the accuracy of our calculation, the additional orbital degeneracy in cylindrical wires merely renormalizes the prefactor of the effective potential $\tilde{V}(q)$, but does not change qualitatively the results we describe. Therefore, the two-dimensional model we propose is a good starting point for modelling Rashba wires at low energies.

Whilst one-dimensional systems are generically very susceptible to disorder, we would like to point out that recent measurements on InSb nanowires \cite{Kammhuber+2016} suggest a mean free path of $1-3 \mu$m is achievable using modern fabrication techniques. This long mean free path puts a low upper-bound on the level of disorder in currently available wires, especially bearing in mind that it exceeds the length of wires used experimentally.

\begin{acknowledgements}
We would like to thank Thomas Sch\"apers, Peter Schmitteckert and Alexander Zyuzin for informative and helpful discussions. TLS \& CJP are supported by the National Research Fund, Luxembourg under grant ATTRACT 7556175. TM is funded by Deutsche Forschungsgemeinschaft through GRK 1621 and SFB 1143. RPT acknowledges financial support from the Swiss National Science Foundation.
\end{acknowledgements}

\bibliography{paper.bbl}
	
\end{document}